A SLIDING WINDOW-BASED ALGORITHM FOR FASTER TRANSFORMATION OF TIME SERIES INTO COMPLEX NETWORKS

Rafael Carmona-Cabezas[1,*], Javier Gómez-Gómez[1], Eduardo Gutiérrez de Ravé[1], Francisco José Jiménez-Hornero[1]

[1]Department of Graphic Engineering and Geomatic, University of Cordoba, Gregor Mendel Building (3rd floor), Campus Rabanales, 14071 Cordoba, Spain

* Corresponding author. e-mail: f12carcr@uco.es

Abbreviations: VG (Visibility Graph), SVG (Sliding Visibility Graph), HVG (Horizontal Visibility Graph), SP (Shortest Path)                                                                                                     1

ABSTRACT

A new alternative method to approximate the *Visibility Graph* (VG) of a time series has been introduced here. It exploits the fact that most of the nodes in the resulting network are not connected to those that are far away from them. That means that the adjacency matrix is almost empty and its non-zero values are close to the main diagonal. This new method is called *Sliding Visibility Graph* (SVG). Numerical tests have been performed for several time series, showing a time efficiency that scales linearly with the size of the series ($O(N)$), in contrast to the original VG that does so quadratically ($O(N^2)$). This fact is noticeably convenient when dealing with very large time series. The results obtained from the SVG of the studied time series have been compared to the exact values of the original VG. As expected, the SVG outcomes converge very rapidly to the desired ones, especially for random and stochastic series. Also, this method can be extended to the analysis of time series that evolve in real time, since it does not require the entire dataset to perform the analysis, but a shorter segment of it. The length segment can remain constant, making possible a simple analysis as the series evolves in time.

**The Visibility Graph is a tool used for the transformation of a time series into a complex network that preserves the features of the original. Nevertheless, as the time series gets larger, the traditional method becomes too slow. In this work, authors propose an alternative way that exploits a property of the network to reduce drastically the time required for very large datasets.**

1. <u>INTRODUCTION</u>

On the current days, investigation on large time series or signals is needed to comprehend a lot of phenomena which appears in many and various fields, from nature to market researches or technological applications. Some examined aspects refer to hidden behaviors or trends and patterns, such as seasonality, which can be expected or not [1–3]. Forecasting, characterization of long-range correlation, chaotic properties and scale invariance are also supplementary spheres where time series were widely studied and promising results were obtained [4–8].

Lately, a new point of view on the study of time series has been born thanks to the transformation of them into complex networks. A recent method named as Visibility Graph algorithm (VG) has been released for this purpose [9]. It is demonstrated that the characteristics of the time series remain in the corresponding complex networks, adding the possibility of analyzing other parameters. Among them, one can find those which measure the centrality of these resulting networks such as the degree distribution [10,11], which will be explained below.

The computation of Visibility Graphs through the basic algorithm requires a time complexity of quadratic order ($O(N^2)$) [12,13]. Another technique which is based on the strategy of "divide and conquer" has been developed in the last years and needs a time computation of order ($O(N \log(N))$ [14].

Additionally, in recent studies some sliding window-based algorithms have been released for very different fields [15–18]. In some cases, these methods have proven to be more efficient in computation time and memory allocation than their respective competitors, especially for large datasets and real-time data. Moreover, some information from datasets is mostly not fully employed, carrying to unnecessary computations.

In this paper, authors have developed a new method which transforms time series into complex networks based in a sliding window algorithm derived from the basic VG principles. It essentially splits the signal into different fixed-size portions, computes the visibility graph basic algorithm for the first window and checks the visibility criterion for each next point. This algorithm has been tested for diverse sizes of windows and several time series as benchmarks. In addition, it has been employed to analyze a real dataset from tropospheric ozone pollution.

The main purpose of this study is to extent the variety of techniques which can be used for converting signals into complex networks, based on the fact that all the information of the time series is not required to build an associated complex network. This can save superfluous computations and time for large datasets by approximating the main characteristics of the corresponding complex networks. The idea is to retrieve acceptable results for small windows sizes compared to the total size of the dataset, with a very narrow confidence interval.

2. MATERIALS AND METHODS

2.1. Visibility graphs

A graph is known to be a set of points (nodes) that are connected by lines called *edges*. In the recent years, a tool that makes possible the transformation from a time series to a graph has gain the interest of the scientific community. This method was introduced by Lacasa et al. [9] and called Visibility Graph (VG), due to its resemblance with the one used to connect points in an Euclidean plane taking into account the possible obstacles [19]. One of its main features is that it inherits many properties from the original series.

In order to build the resulting complex network, it is necessary first to stablish a criterion to determine whether two points in the time series will become connected or not. That condition reads as follow: two arbitrary points from the time series $(t_a, y_a)$ and $(t_b, y_b)$ will have visibility (and will be connected in the graph) if any given point $(t_c, y_c)$ that is located between them $(t_a < t_c < t_b)$ fulfills:

$$y_c < y_a + (y_b - y_a)\frac{t_c - t_a}{t_b - t_a} \qquad (1)$$

In Figure 1, an illustrative time series is depicted in order to show how the points would become connected after applying the VG method. Therefore, the time series is then transformed into a complex network for a posterior analysis. The new network, as it has been said, inherits the complexity of the original signal, meaning that for instance a periodic series would result on a regular graph [9,20].

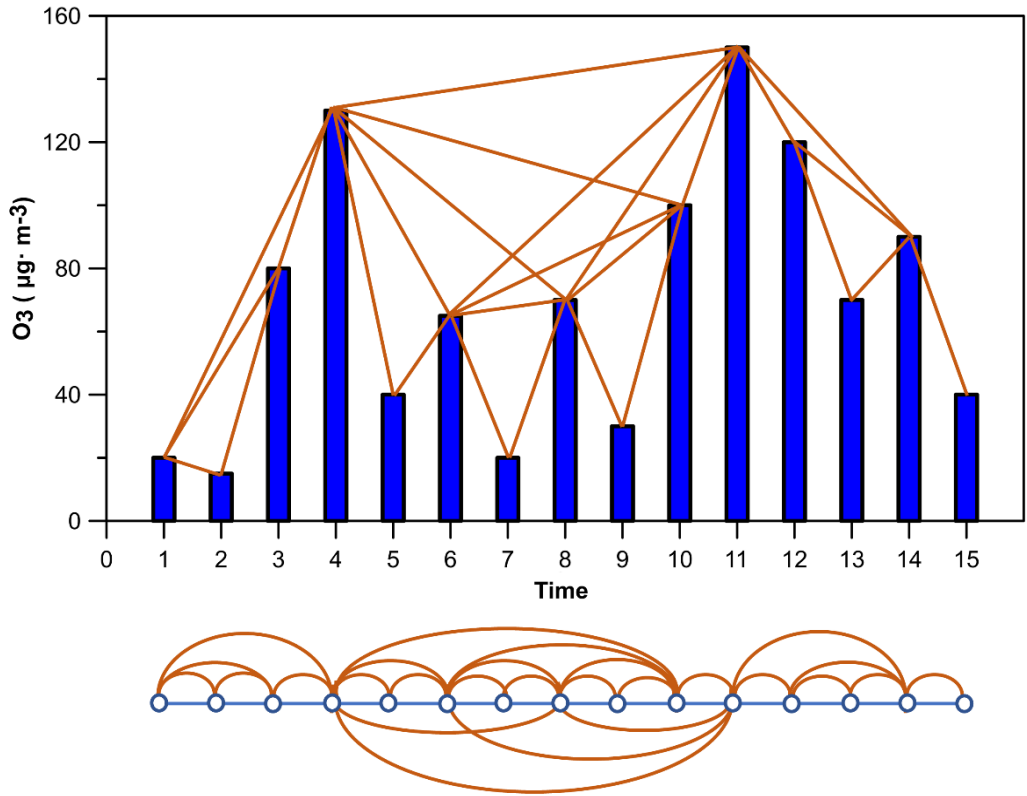

Figure 1: Example time series transformed into a complex network by means of the VG method.

The resulting complex network has several properties, which all the transformed signals have in common, that can be used in order to simplify the algorithm:

- Every node does not have visibility with itself (there are no self-loops in the complex network).

- The resulting graph is undirected, meaning that the visibility between two points is reciprocal and the edges have no direction.

- Every point always has visibility with at least two other points: its closest neighbors, because there are no intermediate nodes to prevent them from fulfilling the condition for visibility.

It can be easily seen in the adjacency matrix $A_{ij}$, which is a representation of the complex network where the elements are $a_{ij} = 1$ if nodes $i$ and $j$ have visibility and 0 if the opposite is true. Taking the previous considerations into account, a VG adjacency matrix has the general form:

$$A_{ij} = \begin{pmatrix} 0 & 1 & \cdots & a_{1,N} \\ 1 & 0 & 1 & \vdots \\ \vdots & 1 & \ddots & 1 \\ a_{N,1} & \cdots & 1 & 0 \end{pmatrix} \quad (2)$$

As a simplification to the VG, another technique was introduced by Luque et al. in 2009[21] to obtain a complex network from a time series with a more restrictive criterion. It was called Horizontal Visibility Graph (HVG). The only difference in the implementation of this method is that the visibility criterion that needs to be fulfilled is:

$$y_c < \min(y_a, y_b) \quad (3)$$

2.2. Sliding Visibility Graph

One of the main problems when it comes to the computation of Visibility Graphs, is the time complexity of the basic algorithm ($O(N^2)$), which is quadratic with the size of the time series, $N$ [12,13]. That is due to the fact that the condition needs to be checked $N(N-1)/2$ times at least in order to evaluate the visibility of every pair of points. When the size of the time series becomes substantially large, the time required to compute the VG rises considerably. Some examples of applications where one can find such great sizes of temporal series are, for instance: datasets that could be studied with a much higher resolution in order to give more light in searched behaviors, such as financial or meteorological time series that are usually regarded from a daily point of view, although much higher resolutions are available [5,22]. Also, there are time series which have a huge amount of points that need to be split for the convenience of the computation. One example can be found in time series of musical compositions, where only a small piece is usually taken; while it would be of interest to analyze the whole series [23,24].

Here, authors propose an alternative to the original VG that has a time complexity of $O(N)$, which ends up being very convenient for computation of very large time series. The name given to this new algorithm is *Sliding Visibility Graph (SVG)* and the idea behind lies in the fact that most of the adjacency matrix of such complex networks are almost empty as one goes further from the main diagonal. That means that the vast majority of the iterations are considering pairs of nodes that will not have visibility.

The basic principle of this method is to take a set of points of length $W$ (which we will refer to as "window" in the text) to check the visibility criterion inside of it and then start moving that window towards the end of the series. In practice, it is desirable to initialize the routine with a regular VG for the first $W$ points, in order to have all the information from those first $W$ points. The main advantage of this technique is that each time that the window is displaced, it is not necessary to compute again the visibility of all the points included within. Since most of them were already computed in the previous iteration, only the new point needs to check the criterion with the prior $W$ ones ($W$ times). It is easy to check that the number of total iterations $n_i$ needed will be as shown in equation 4.

$$n_i = (N - W)W + W(W - 1)/2 \qquad (4)$$

The last term of the sum comes from the initial iterations required for the computation of the first VG with the length of the chosen window. It is derived from the number of iterations needed for the basic VG algorithm, as explained before. This term will be negligible for $W \ll N$, while it will be the responsible for most of the iterations as $W \to N$.

In Figure 2, the exposed procedure can be graphically observed, for the sake of clarity. Figure 2a) illustrates the cited initial iteration that requires a regular VG.

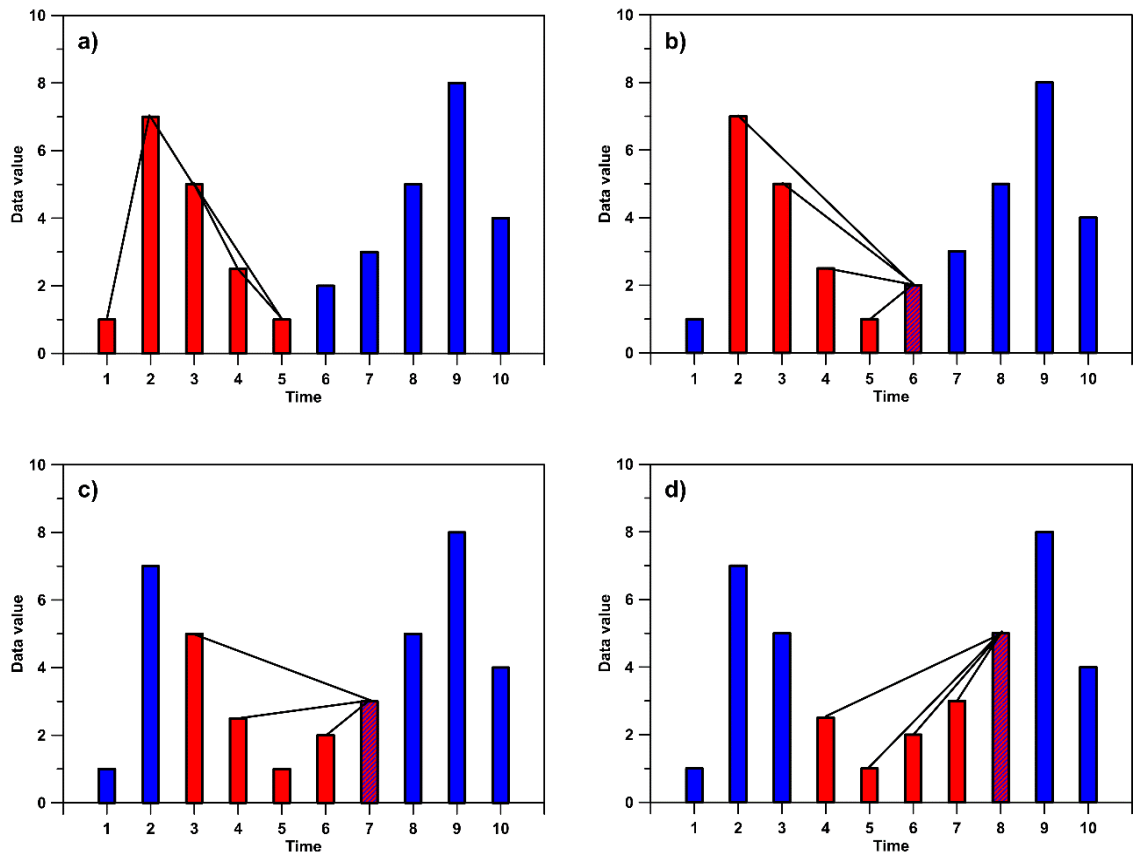

Figure 2: Illustration of some example iterations of the SVG method. The plot a) shows the first basic VG that is performed, while b), c) and d) are the posterior three iterations.

Obviously, this method constitutes an approximation to the original VG as long as the size of the window is smaller than the total one of the time series. Nevertheless, due to the difficulty for two points to become connected nodes as the distances between them gets higher, the results derived converge very rapidly to the original ones, as will be shown in later sections. When the size of the window and the time series are equal, it is trivial that both SVG and VG are equivalent.

2.3. <u>Degree centrality</u>

When trying to retrieve information from a given complex network, one of the most commonly used approaches is discerning which of them are the most important nodes in the system. To this purpose, centrality measures come usually in handy. This concept was initially applied to the study of social networks and later transferred to other fields of knowledge [25–27]. This work has been focused on one of them: the degree centrality measure, which will be explained next.

The degree of a node ($k_i$) in a graph is the number of other nodes which are connected to it ($k_i = \sum_j a_{ij}$) and in this case, the one that it has visibility with. For instance, in Figure 1, the degree of the three first nodes are $k = 3$, $k = 2$ and $k = 3$, respectively. After computing the degree of the whole network, it is possible to obtain the probability that corresponds to each value, by simply counting how many times each possible one is repeated. From there, the degree probability distribution of the sample $P(k)$ can be obtained.

One possible way to describe the nature of the time series is to analyze the degree distribution that is built from the VG as previous works have shown [9,10,28], being able to distinguish between fractal, random or periodic signals, for instance. As some former studies explain [13,20], time series which have VGs whose degree distributions can be fitted to a power law $P(k) \propto k^{-\gamma}$ correspond to scale free ones; which can be explained due to the effect known as *hub* repulsion [29]. The term *hub* refers to a node with an unlikely high number of links with respect to the rest of the network (the *hubs* have the highest degrees). The right tail of each degree distribution, governed by those *hubs*, can be represented in a log-log plot and fitted by a simple linear regression when the series obeys a power law. The slope obtained by this regression provides an interesting parameter, the so-called $\gamma$ exponent, which has already been used in some works [20,28] to retrieve some useful information about the signal.

### 2.4. Complementary centrality parameters

There are several other centrality measurements that have been considered to check the correct suitability of the proposed method in order to describe the same complex network. One of them is the so-called shortest path (SP), which can be understood as the minimum number of edges that connects two arbitrary points. More precisely, two distant nodes $(i,j)$ will have different number of edges and paths (distinct configurations of edges that link both) between them, but there will be some of these paths where the number of edges will be minimum (minimal path is degenerated); this quantity is named as the SP.

After introducing this parameter, one can define the closeness centrality. If all pairs of nodes are considered, it is possible to obtain a matrix, which is called distance matrix $D$. In this matrix each element $d_{i,j}$ refers to the SP from node $i$ to $j$. Diagonal elements are set to zero. For an undirected graph, this matrix is symmetric, as in the adjacency matrix case. It is possible to define the named closeness centrality of a node $i$ as the inverse of the sum of distances from this node to the others, i.e.:

$$c_i = \frac{1}{\sum_{j=1}^{N} d_{i,j}} \tag{5}$$

Where $d_{i,j}$ is the element $(i,j)$ from the corresponding distance matrix of the graph.

Finally, betweenness parameter has been added to these, which is a measurement of how a node is between many others. That is, how much a node is passed through by shortest paths of other pairs of nodes. Therefore, this quantity can be defined for a node $i$ as:

$$b_i = \sum_{\substack{j=1 \\ j \neq i}}^{N} \sum_{\substack{k=1 \\ k \neq i,j}}^{N} \frac{n_{jk}(i)}{n_{jk}} \tag{6}$$

Where $n_{jk}$ is the number of SP's from $j$ to $k$ (remember that these paths can be degenerated), whereas $n_{jk}(i)$ is the number of SP's that contains the node $i$.

3. <u>RESULTS AND DISCUSSION</u>

3.1. <u>Performance tests</u>

In this section, the proposed algorithm is tested by using several time series with different nature. All the studies in this work have been performed using MATLAB 2018b on an Intel(R)Core(TM) i7-8700 CPU @ 3.20 GHz, with a RAM of 8 GB and the O.S. Windows 10 Education x64.

In Figure *3* the time series used for testing the algorithm are shown in the upper part. The first one (a.1.) corresponds to a random series (from 0 to 1) with $N = 5 \cdot 10^5$; the second one (b.1.), to a Brownian motion time series with a Hurst exponent of $H = 0.5$ and $N = 2 \cdot 10^4$; while the last one (c.1.) is a stochastic one of hourly ozone concentrations recorded in the southern part of the Iberian Peninsula in 2013 (nearly $N \approx 8500$). On the other hand, the bottom plots show the algorithm computation times on each time series, for both SVG and original VG. It must be noticed the fact that the window size (W) in the x-axis of the last mentioned plots are normalized to the total size (N) of each example series. It has been done in order to compare the different tests, since all the chosen values of $N$ are unequal.

As it can be regarded in the lower part of Figure 3, the behavior of the computation time with $W$ is parabolic, having its maximum when $N = W$. To check that quadratic trend, a second-degree polynomial fit was performed to the curves and in all the cases the Pearson coefficient was greater than 0.9999. At the maximum of the curve, the computation times of both methods are the same since they are equivalent when those two quantities are equal (as explained before). Before that maximum is reached, the computation times of the SVG are always lower than those of the original VG. That is because, as it can be seen in equation 4, the number of iterations of SVG can never be greater than the amount needed for the VG, since by definition $W \leq N$.

In the case of the ozone concentrations, the measured computation times have a bigger noise due to the lower size of the time series, for which the computation times are considerably reduced, compared to the other two.

Once the scaling with the size of the window $W$ has been discussed, the equivalent analysis with $N$ is necessary as well. Indeed, it is where one of the main advantages of SVG with respect to the original can be observed, since the scaling changes from $O(N^2)$ in VG to $O(N)$ in the new proposed one.

In Figure 4, the computation time required to perform the SVG method is plotted against the size of the analyzed time series. The same series as in the previous case were used, but in this case varying their size from 10 to the original one. The result for all of them is the same: the original VG has a time efficiency that scales quadratically with the size of the time series ($O(N^2)$), while the SVG behavior with respect to $N$, is always linear ($O(N)$). In these plots, the absolute values of $W$ have been used instead of the ones relative to the value of $N$ as in Figure 3, because its value is now variable and it could be misleading. Instead, three different lengths of window have been shown (10, 500 and 1000). As it is clearly seen, the larger is the window size, the greater the slope of the computation time over $N$. The same tests were performed for higher values of $W$, obtaining the same linear behaviors, despite having a shorter curve (because $N$ can only be evaluated from a $W$ size onwards). These results are in accordance to what was expected from equation 4.

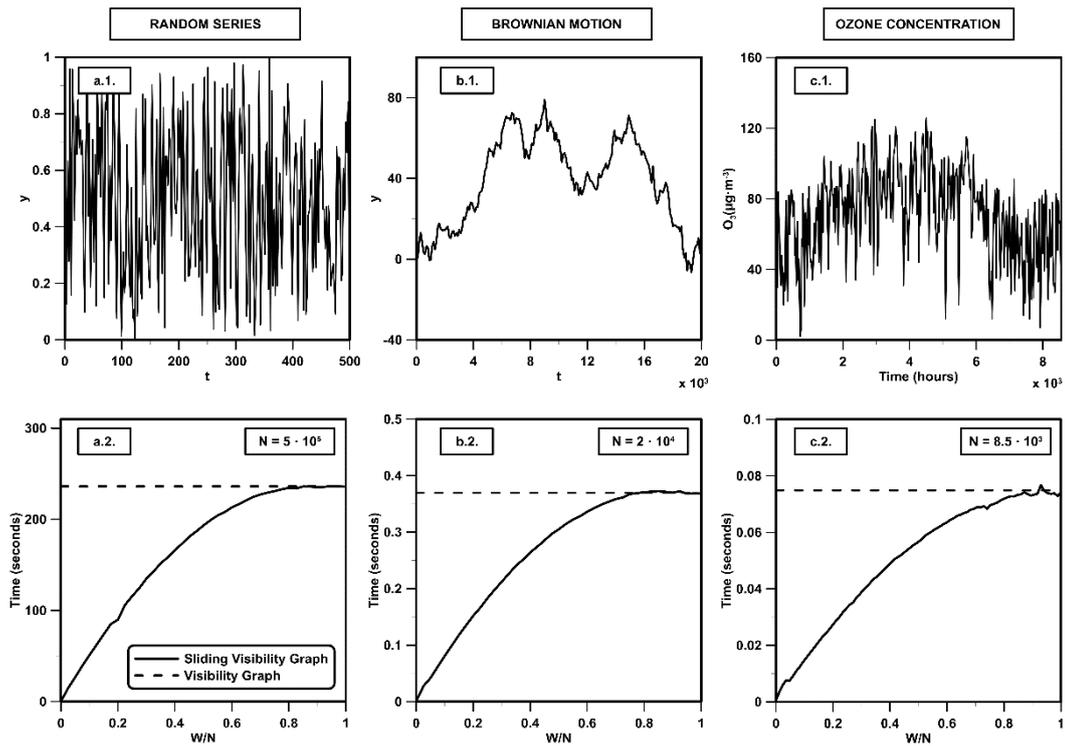

Figure 3: Time series used for the analysis (up) and computation times versus window size (down). Dashed line indicates the time required for the basic VG method. In a.1., only a portion of the total series is shown, for clarity reasons.

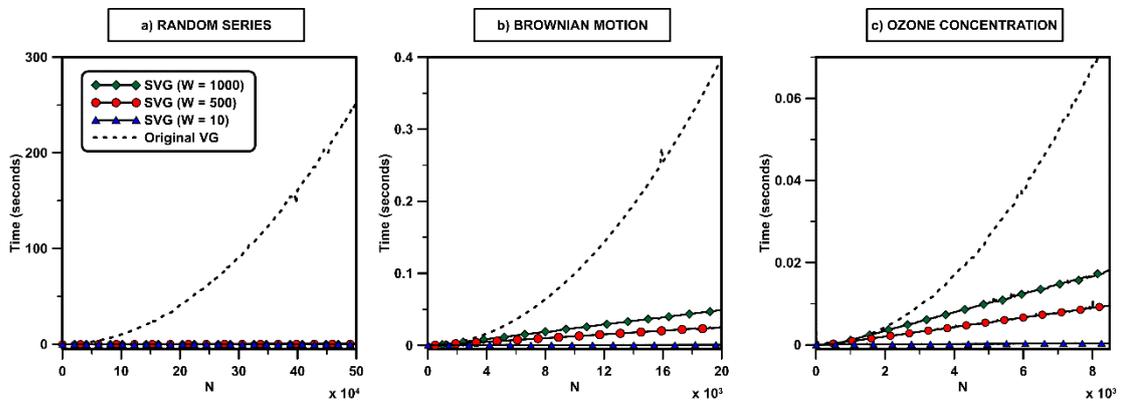

Figure 4: Computation time versus the size of the time series for several values of window size.

3.2. Degree distribution results

After the computation efficiency of this method has been tested, the next step is to check whether the results of the SVG merge the ones of the VG and when this approximation is valid. To see this, in Figure 5 the degree probability distributions of the three series used before are shown. There, the distribution of each series is computed using the regular VG, the HVG and then with the SVG, with several increasing sizes of window. The same relative values of $W/N$ are used in the three cases for clarity reasons. The series sizes $N$ in all of them is fixed to the original values used for discussion in Figure 3, hence the study conditions remain the same as in the previous section. A log-log plot is used, because one of the main uses of this degree probability distribution is the acquisition of the $\gamma$ exponent. That is done by fitting the linear tail in the last part for scale free networks using the logarithmic scale, as explained. Some works have already shown that those series degree distributions can be fitted to a power law $P(k) \propto k^{-\gamma}$ [10,20]. Therefore, in Figure 5 b) and c) it was interesting to check whether this linear behavior of the tail is hold with the SVG as well.

For the three series, it can be observed that the degree distribution differs evidently from the original for very low values of W, especially for the Brownian motion case and the ozone time series. In those cases, it is seen that the largest degrees of the distribution (hubs) are always sensibly lower than the actual hubs degrees in the original VG. That is due to the fact that the chosen window ($W = N/500$ in the plots) is still too narrow to allow hubs to check the visibility with most of the nodes they are connected to. Thus, the biggest values of k for those windows will have a limit given by twice the size of the window ($2W$), since a node $i$ will check its visibility with others only from $i - W$ to $i + W$. That is in accordance with the results in Figure 5 b), where for $W = 40$ ($N = 2 \cdot 10^4$) there are no values of $k$ higher than 80 ($2W$), for instance. Hence this can be used as criterion for choosing a minimum value of $W$ that allows the convergence $P(k)$.

On the other hand, when $W$ increases, the degree distributions of SVGs start to resemble the original ones. So that at a value of $W = N/100$, they are almost overlapped and big changes on the length of the window (for instance $W = N/10$) give almost equivalent results on the random and ozone series. In the case of the Brownian motion, this convergence is slower, being necessary larger sizes of window. The reason behind this phenomenon is the roughness of the series that depends on the Hurst exponent [30]. Nevertheless, it must be pointed out that the computation time remains lower than in the original case no matter the value of $W$, as discussed previously. This saturation can be seen more in detail in Figure 6. In every case, networks from the HVG in Figure *5* give distributions that are very different from the original VG, having SVG a better approximation of the distribution for all the selected windows.

To test the convergence of some actual numerical values rather than the shape of the curves, the average degree $\bar{k}$ of each distribution has been computed (Figure 6) as well as the γ exponent from the linear regression (Figure 7) and other parameters described before. Now, it is only shown for the case of the Brownian motion and ozone concentration time series, since the random series does not have a power-law behavior and for the value of $\bar{k}$ will converge much faster than the others, as can be deducted from Figure 5. Hence, the two series that fit the original case worse are looked at in more detail.

Regarding the top part of Figure 6, the value of $\bar{k}$ obtained through the SVG against the VG and HVG can be seen. As it was expected from the previous commented figure, the merging in the case of the Brownian series is slower than in the case of the stochastic ozone one. Nonetheless, its shape converges faster than the number of iterations (quadratic) even in the Brownian case, as discussed in relation to the time performance of the method (see Figure 3).

On the other hand, when the size of the series itself is increased, the $\bar{k}$ obtained with the SVG and a fixed value of $W$ converges as well. That is shown in the lower part of Figure 6. In this case, for the ozone time series a window of $W = 0.1 \cdot N_{max}$ is enough for the convergence in all the sizes tested; while for the case of the Brownian motion, it is necessary to reach a value of $W = 0.25 \cdot N_{max}$. On the contrary, it is seen in this same figure how the values of the HVG are always lower than those of VG, being the SVG ones greater even for very small windows. This fact could be anticipated by looking at Figure 5.

To provide an adequate criterion to decide a proper optimal window, authors propose a technique that will be further explained. After several test with time series of different natures and lengths, it has been determined that the SVG results always saturate when they approach the value of the original VG. This leads to the difference between results from consecutive windows decaying to zero. This can be used to determine a saturation point by computing the slope of the curve for each window. When this slope is considered to be close enough to zero, the optimal window is taken. The criterion for this consideration is that the slope between the two last points ($W_i$ and $W_{i-1}$) falls below a percentage of a characteristic value automatically set for a given time series ($\bar{k}_\iota/N$). Authors have detected that a 5% of this value is suitable to an optimal window. Results for the average degree computed can be seen in Table 1, where the ratio between the optimal window and the total size ($W_{Opt}/N$) is shown in the second column, the ratio of the required times in the third one ($t_{SVG}/t_{VG}$), and the relative error ($\varepsilon_k$) from the computation of the average degree in the last one. Since all of them are relative values, they are expressed as percentages for the sake of clarity.

| TIME SERIES | $W_{Opt}/N$ | $t_{SVG}/t_{VG}$ | $\varepsilon_k$ |
|:---:|:---:|:---:|:---:|
| Random | 0.8 % | 1.7 % | 0.03 % |
| Brownian ($H = 0.5$) | 27.3 % | 45.3 % | 1.08 % |
| Annual ozone concentration | 5.5 % | 10.2 % | 1.28 % |

Table 1: Results obtained from the optimal window criterion for three time series.

In order to test the applicability of the SVG, four more different parameters (γ exponent, SP, betweenness and closeness) are going to be described in the following part of the presented document. Those have been previously employed to study both visibility graphs and complex networks [22,31]. The results in Figure 7 and Figure 8 show the values of these quantities after applying the SVG for several windows (as it was done before with the degree) and their comparison with the VG and HVG methods.

The results depicted in Figure 7 show a similar behavior to those in the upper part of Figure 6, but this time with the slope of the distribution tail (γ exponent) and the SP. The γ exponent of both time series (upper part) tends to the value obtained through the basic VG as the size of the window increases. Again, the Brownian series needs a larger size of window to have a negligible error from the SVG, due to its greater roughness, as discussed. Nevertheless, a good fit is already achieved for windows of around 25% of the entire size of the series, meaning that the information of the hubs in the system can be retrieved by using this method. Regarding the SP (bottom part of the same figure), a fast convergence is as well observed. For very low values of W, the SP is higher in comparison to VG, because such small window does not allow fast connections between distant time points that would lead to the real SP, and therefore the distances are greater. As this window is increased, those fast links are possible, and the SP is decreased.

Finally, Figure *8* shows now the two next computed parameters: betweenness (up) and closeness (bottom). Both are in accordance with the results discussed previously. The reason for the convergence can be understood in the same way as what was explained above for the case of SP.

Once again, outcomes from HVG differ substantially from the VG ones, being the SVG results closer to them. This fact is even more pronounced in the case of the Brownian motion time series (in both last figures). Authors attribute this effect to its higher irregularity, that leads to a worse overall performance of the HVG algorithm. Since the HVG algorithm excludes nodes once a higher value than the initial one is found, it can omit important nodes that VG and SVG (with a proper window) would not. That, in the sense of connections between very distant points in the time series, is very important for the computation of the three last parameters mentioned. For that reason, the approximation of the HVG can even differ several orders of magnitude from both VG and SVG ones (see Figure *8*b1).

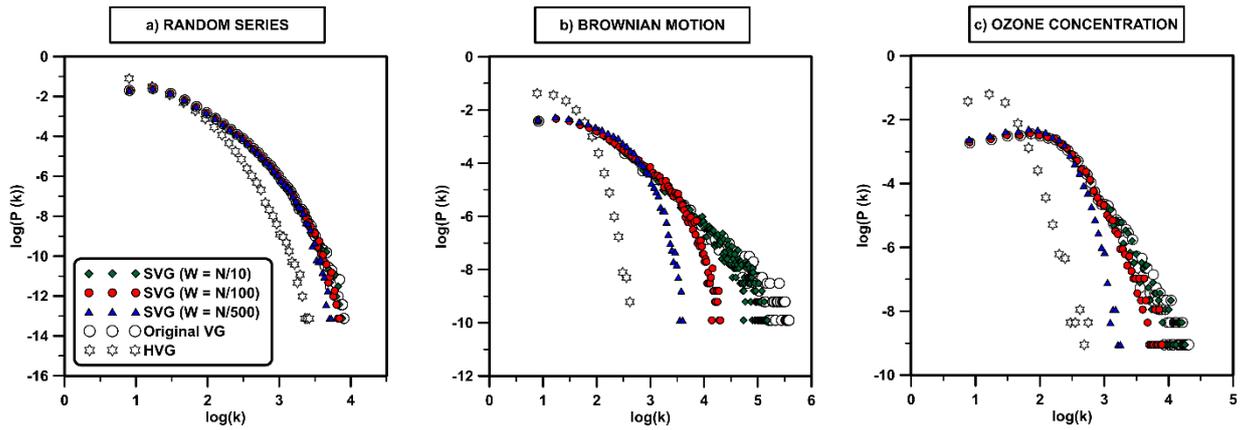

Figure 5: Degree probability distributions of the networks obtained by the VG and SVG are shown here. Three window sizes from $W = N/1000$ to $N/10$ are used for comparison in the SVG.

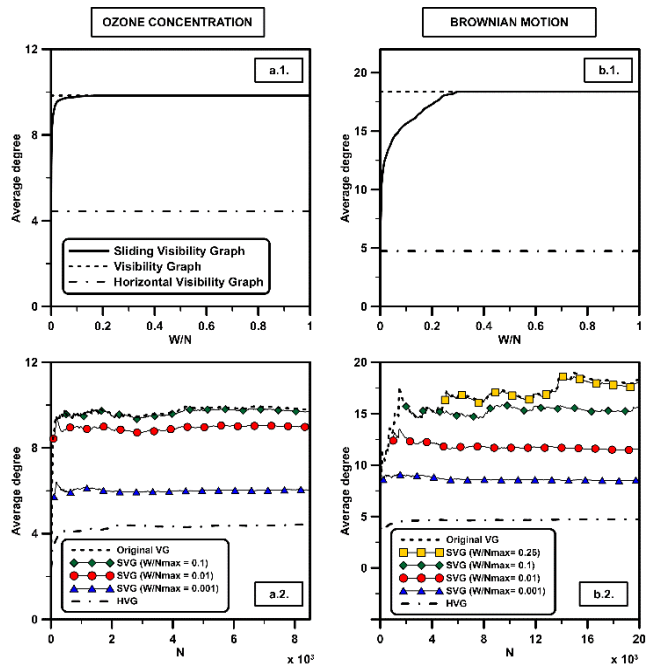

Figure 6: Average degree obtained through VG and SVG versus the window size (up) and versus the total series size (down).

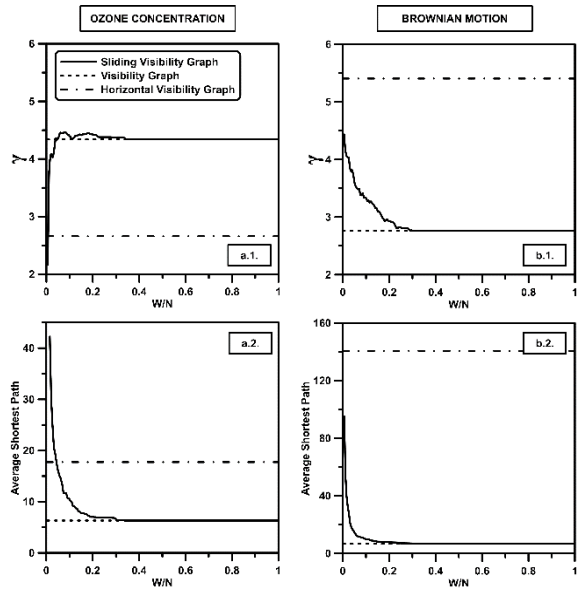

Figure 7: γ-exponent (up) and average SP (bottom) computed by using the original VG method (dashed line), the HVG (dashed-dotted line) and the SVG for different values of window length.

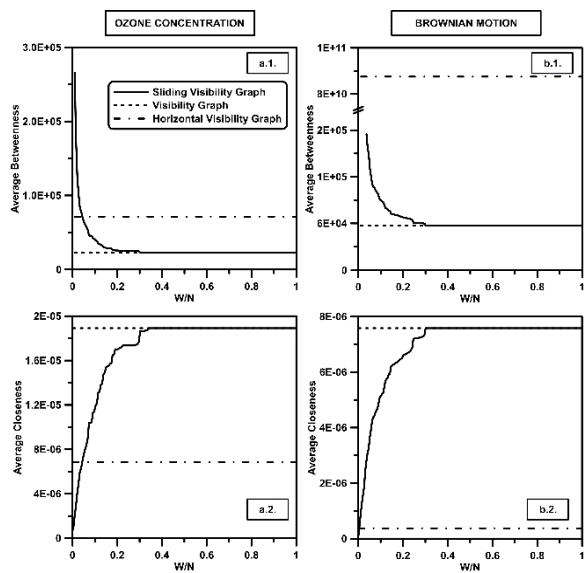

Figure 8: Average betweenness (up) and closeness (bottom) computed by using the original VG method (dashed line), the HVG (dashed-dotted line) and the SVG for different values of window length.

4. CONCLUSIONS

Throughout this work, a new method to compute the Visibility Graph out of a time series has been introduced and tested. This new approach, the Sliding Visibility Graph, approximates the original VG and has its basis on the fact that the adjacency matrix of the network built is almost empty and all the values tend to be as close as possible from the main diagonal.

The performance was tested on several time series with different nature (random, fractional Brownian motion and real stochastic measurements) and the results show that the time efficiency has a parabolic trend with respect to the window size $W$ and linear with the size of the series $N$. By definition, the number of iterations needed by the SVG is going to be always lower than those of the original VG, leading to a faster performance for every time series. In the limit case of $W = N$, SVG converges to the basic VG, bringing obviously to the same time performance.

When it comes to the results obtained from this alternative method, as expected from the properties of the adjacency matrix, the SVG outcomes rapidly converge to the ones obtained by the VG for low sizes of the window. This has been demonstrated for several distinct parameters. The main advantage of this is that once a proper window is chosen for a kind of time series, the correctness of the approximation appears to hold for larger sizes of the series, making it very convenient for great data series. Authors propose a technique that can be automated in order to find the optimal windows and has been tested to provide satisfactory results. It has been compared to HVG results as well, showing that SVG outperforms it for almost every chosen window.

To conclude, it must be pointed out that SVG constitutes an alternative approximation to the widely used VG and HVG that could have a big potential for several cases. In particular, authors would like to mark two main scenarios: i) very large time series, where the computation times would be huge and ii) real time analysis, where the size of the system gets larger with every measurement. This will be studied in future works, in order to check whether the parameters derived from these complex networks can be used to predict changes in the behavior of a temporal variable, for instance.

5. ACKNOWLEDGEMENTS

The FLAE approach for the sequence of authors is applied in this work. Authors gratefully acknowledge the support of the Andalusian Research Plan Group TEP-957 and the XXIII research program (2018) of the University of Cordoba. R. Carmona-Cabezas truly thanks the backing of the "Programa de Empleo Joven" (European Regional Development Fund / Andalusia Regional Government).

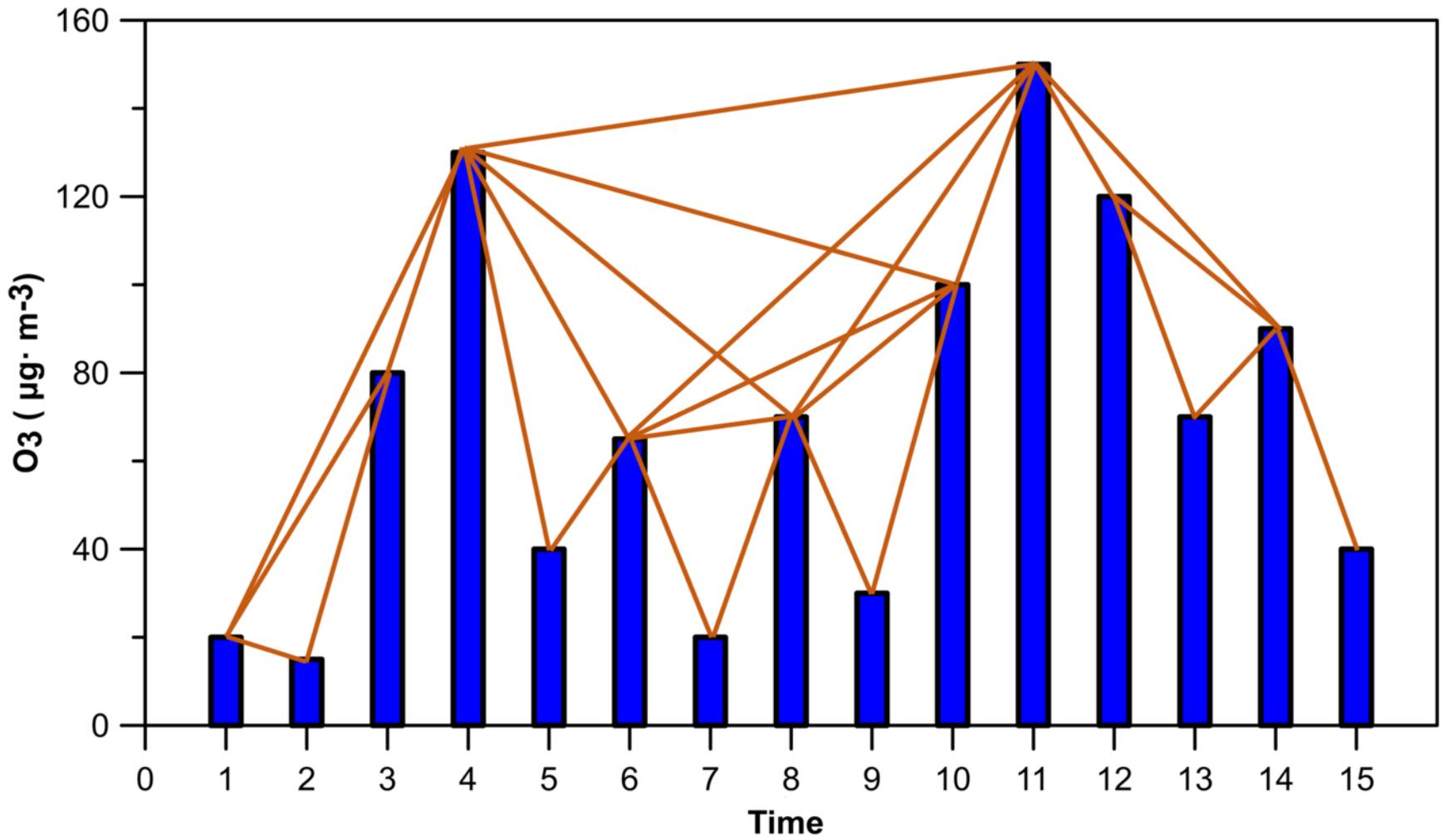

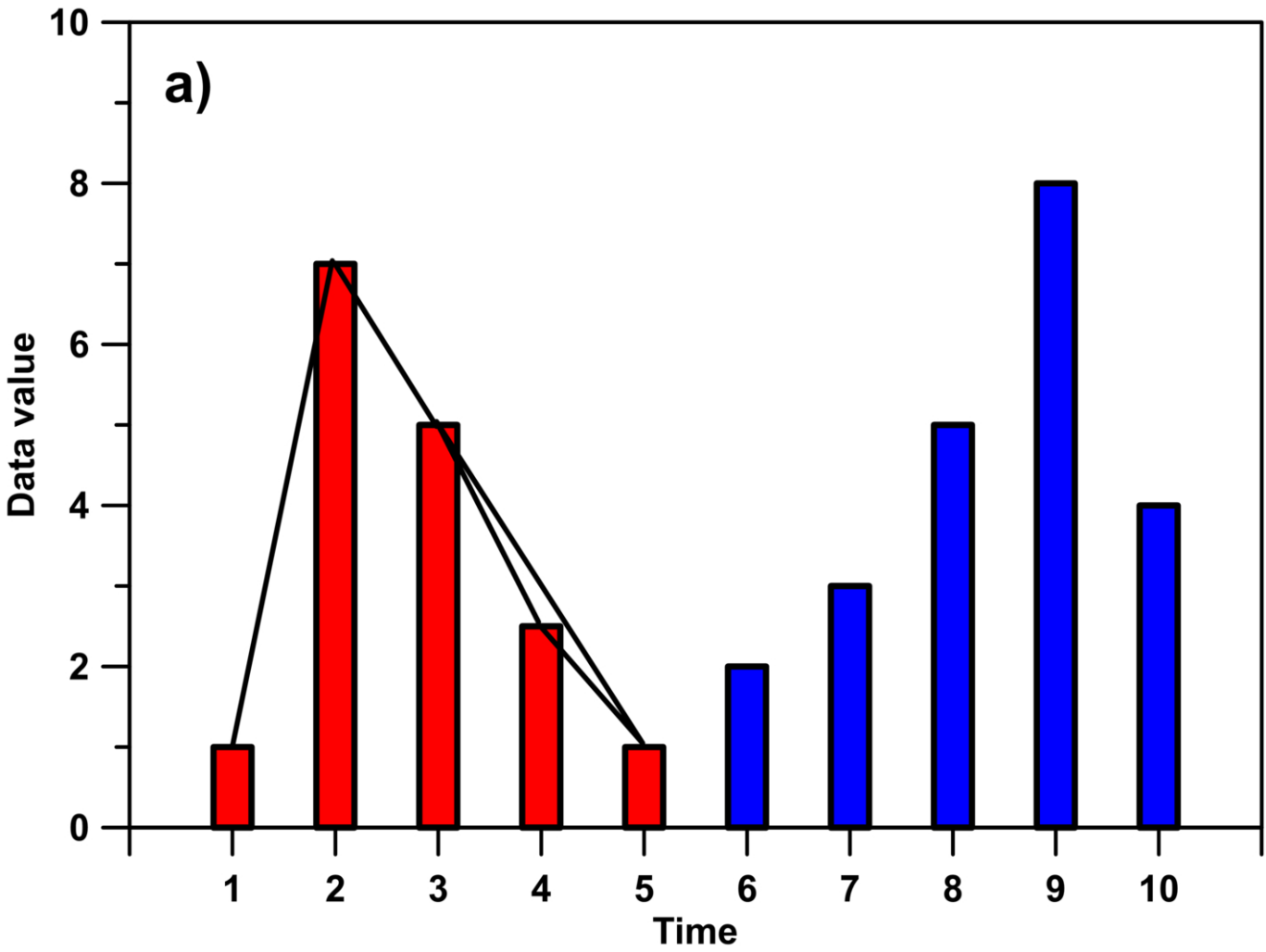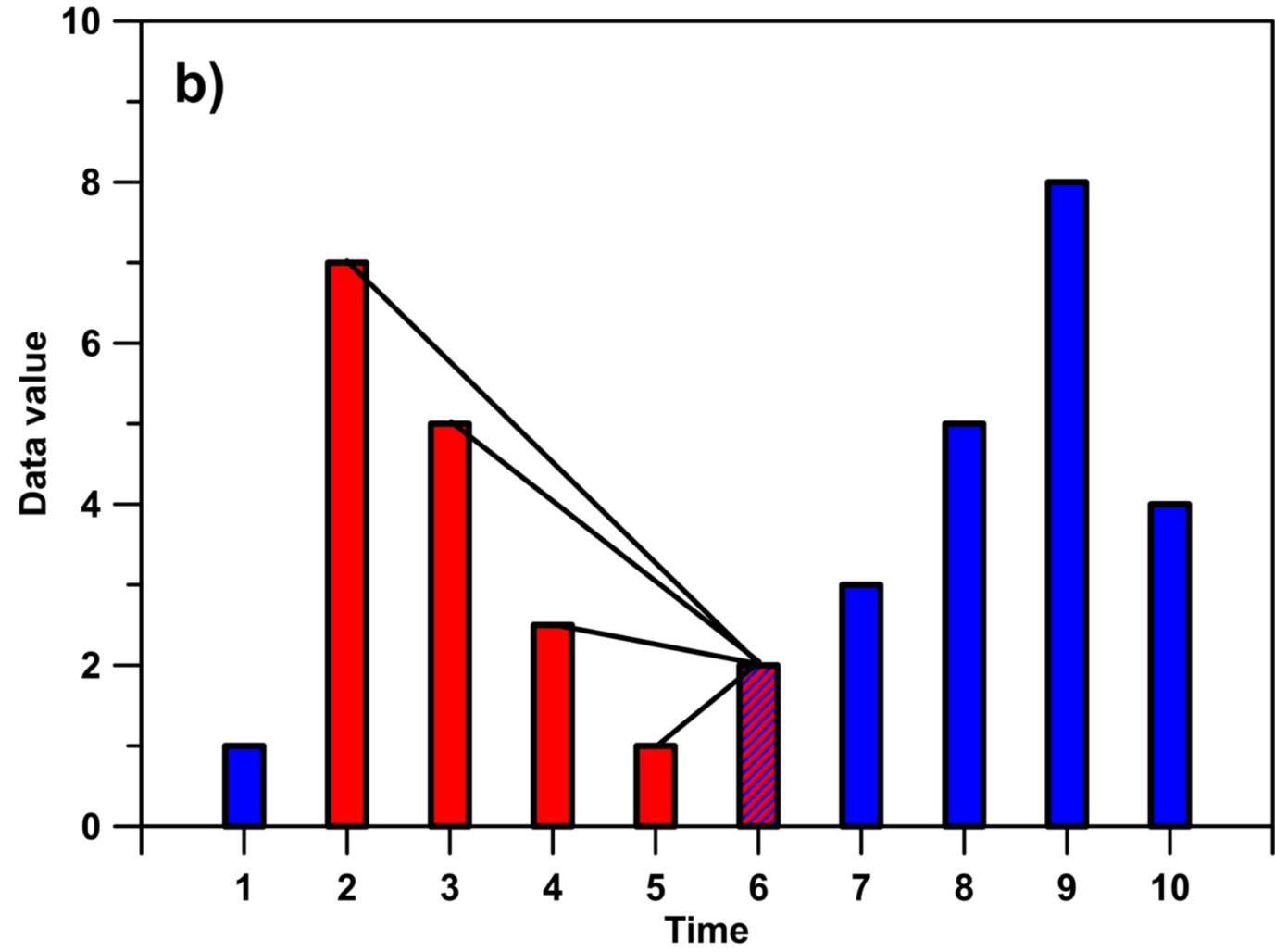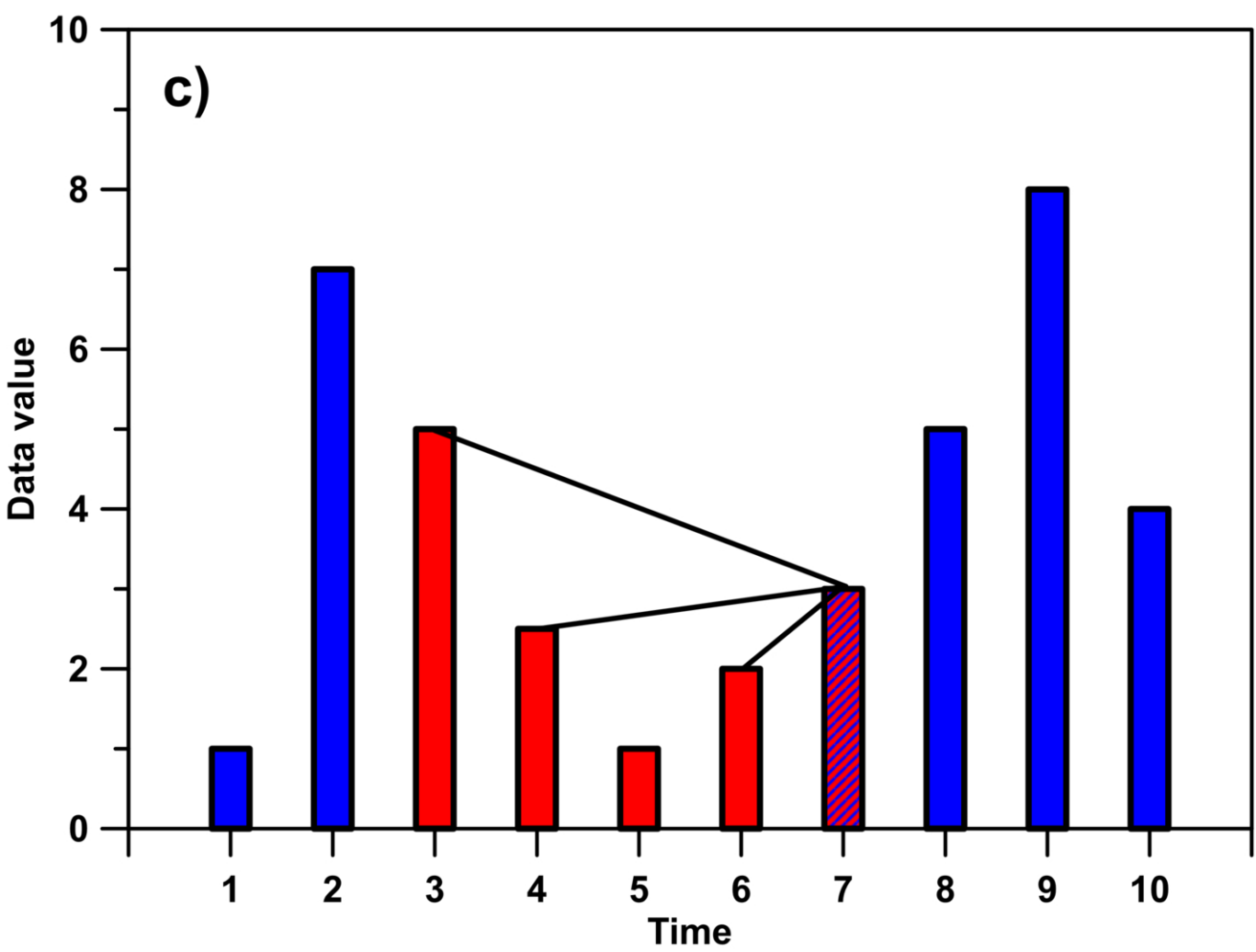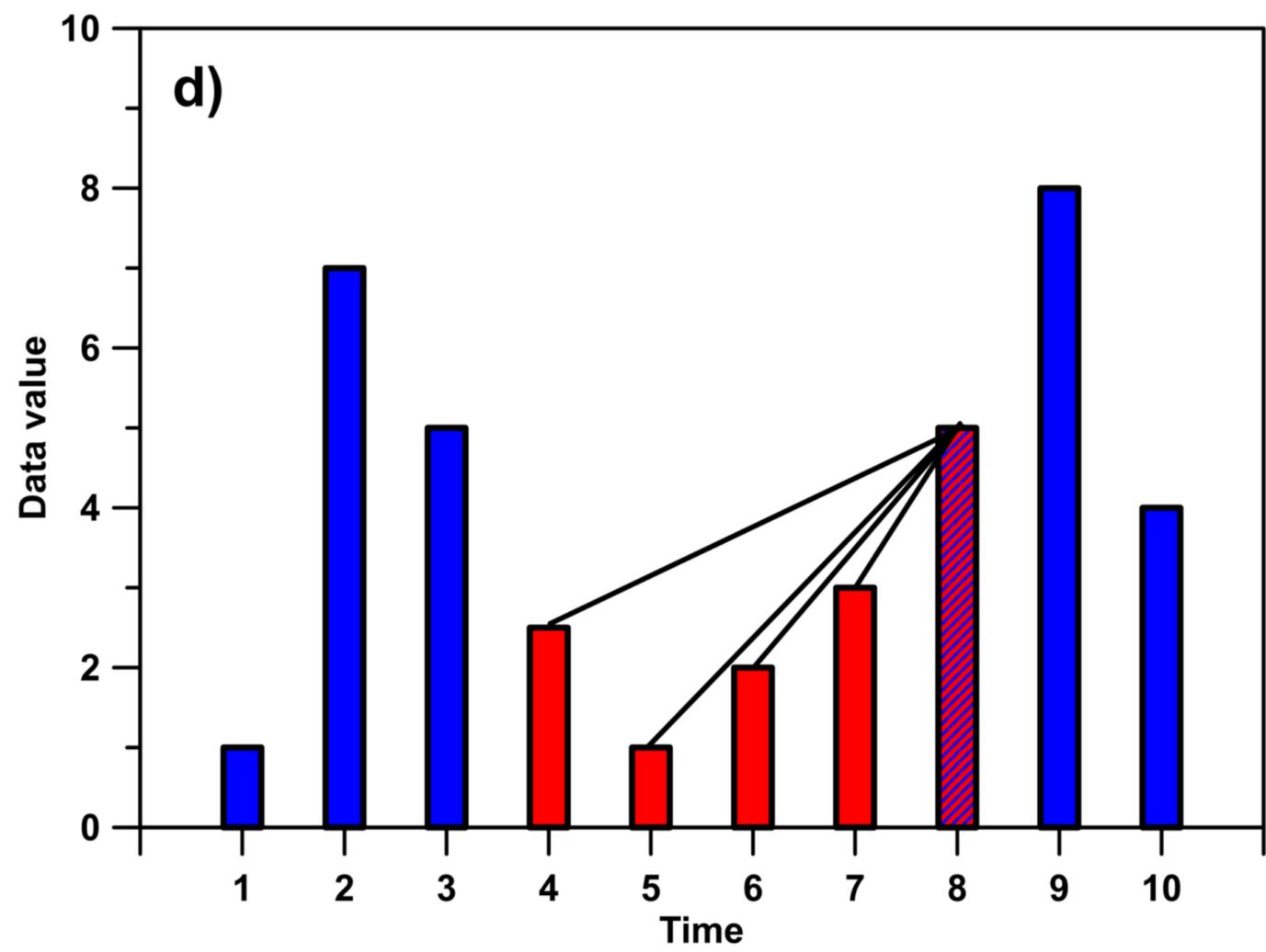

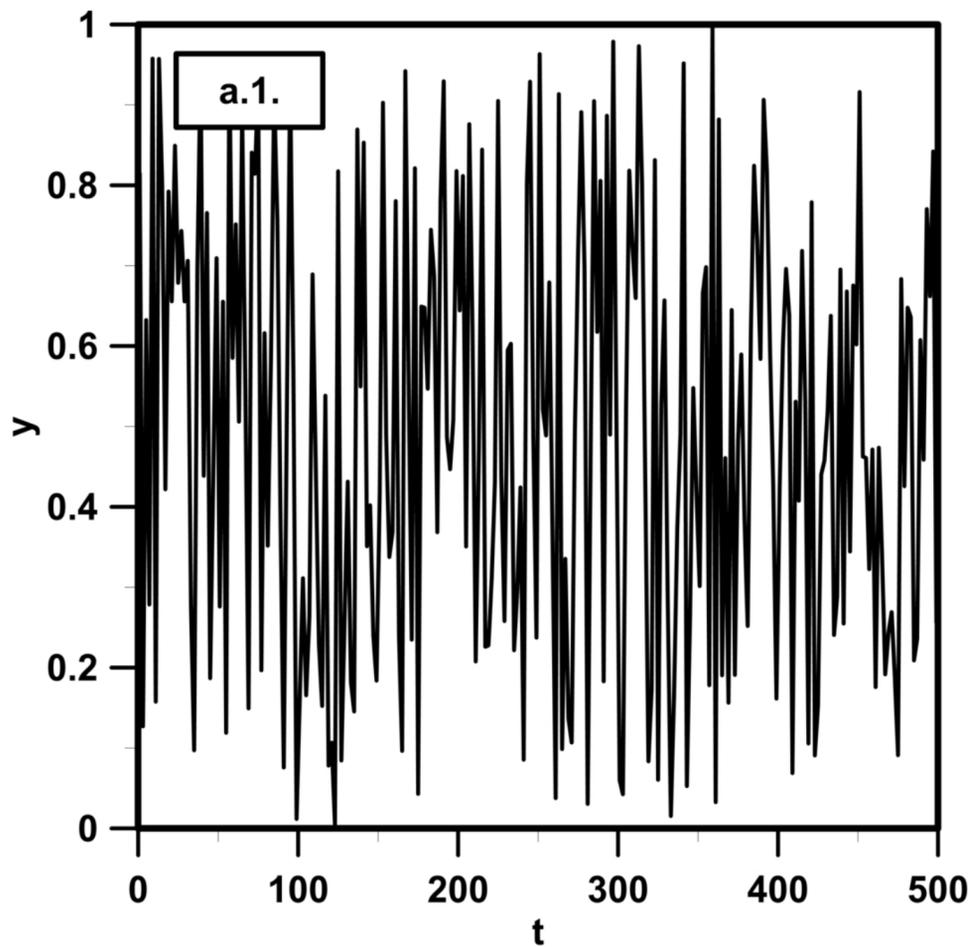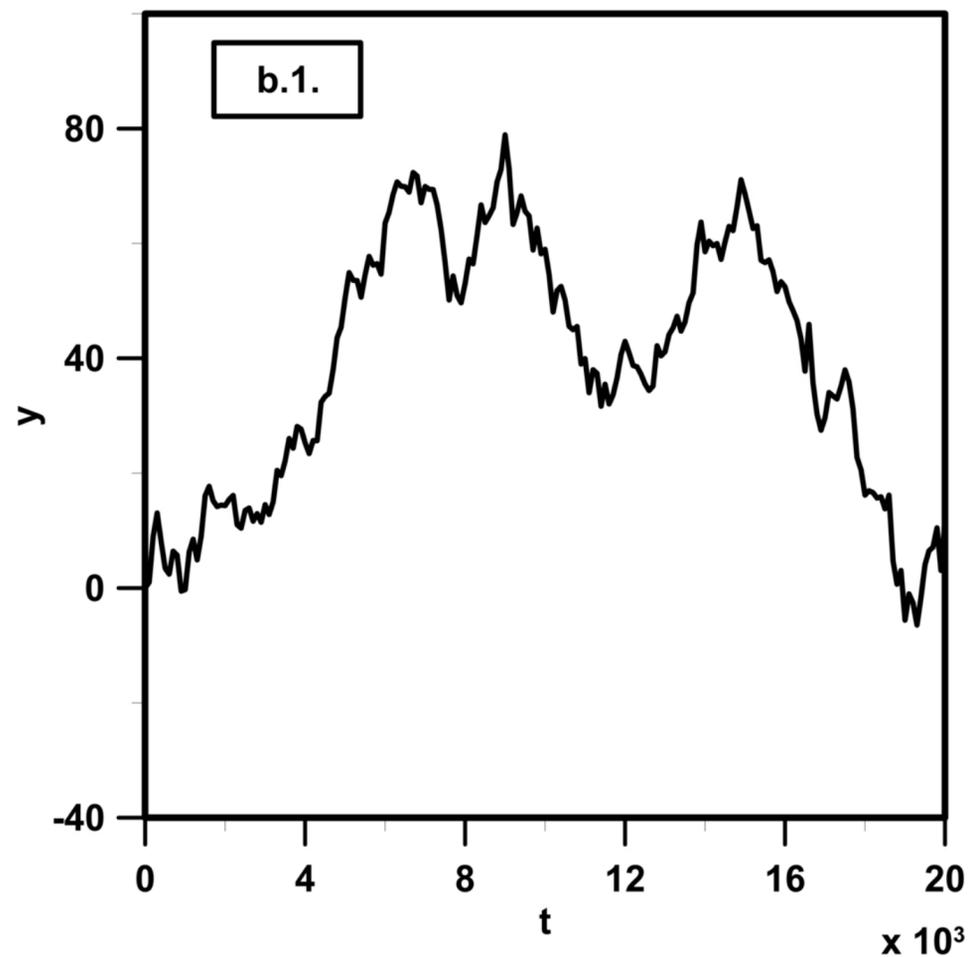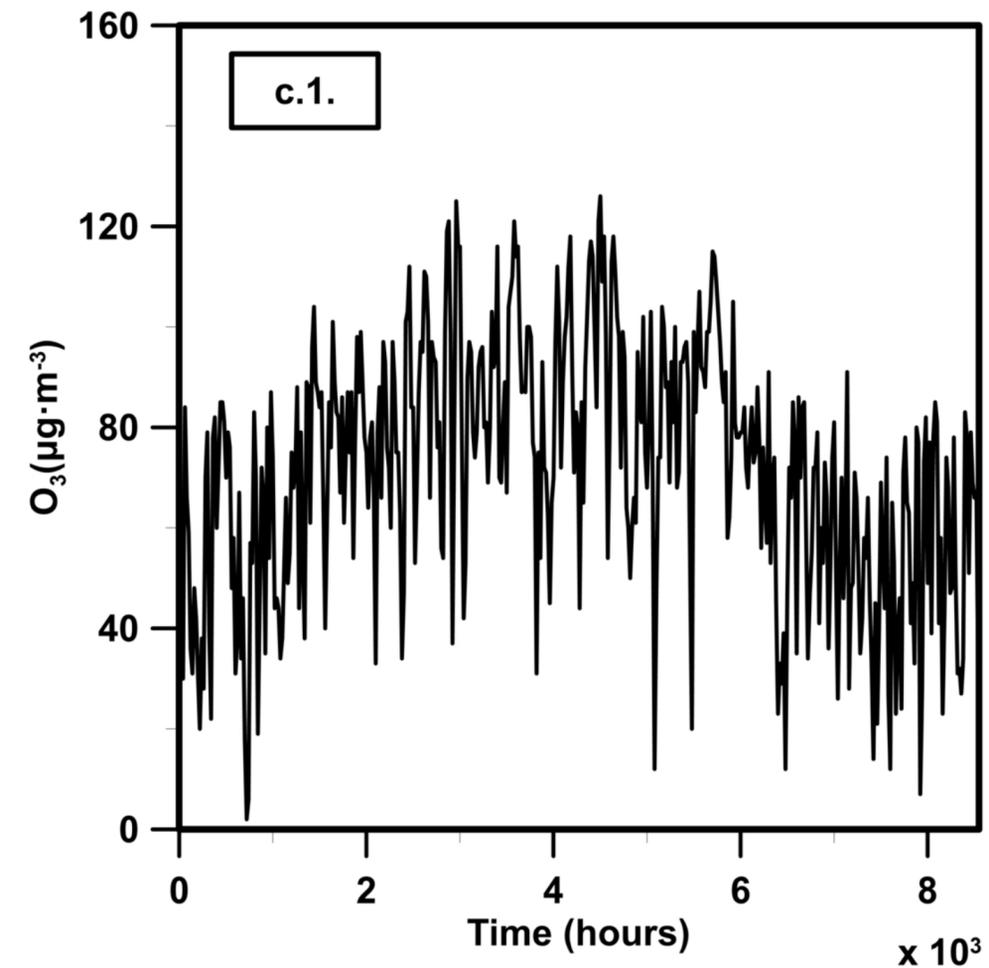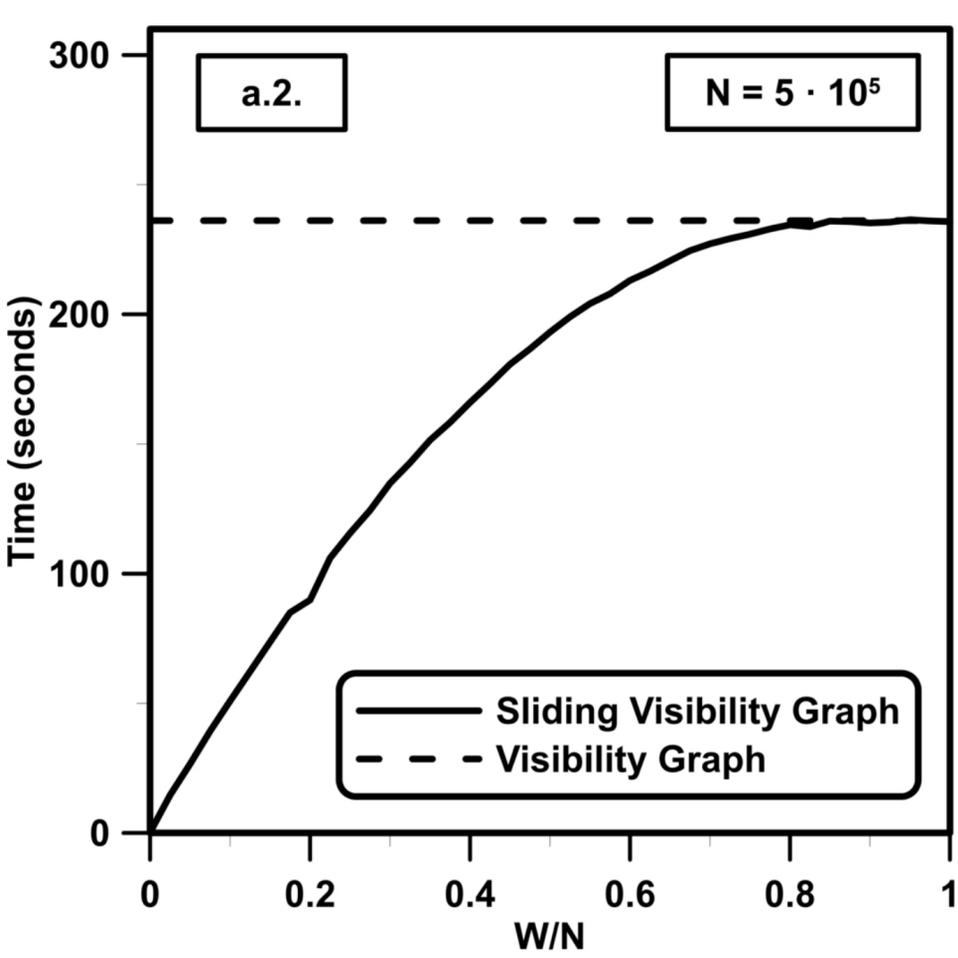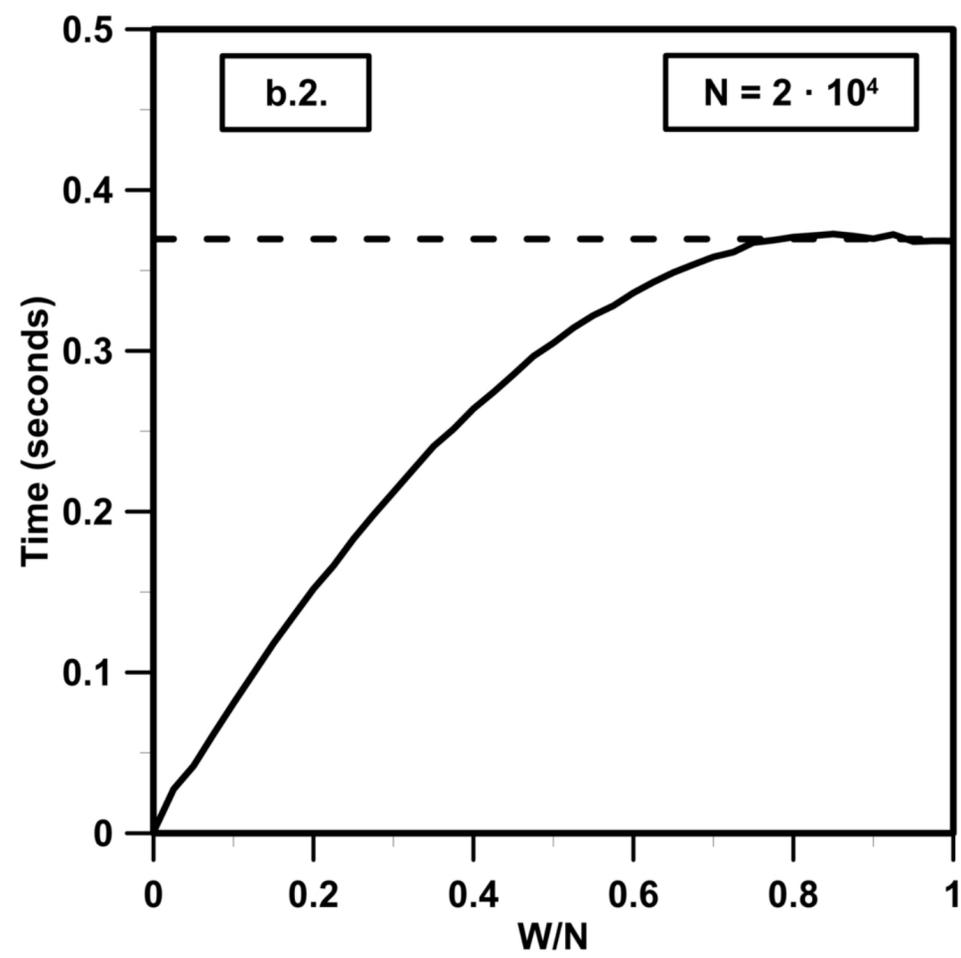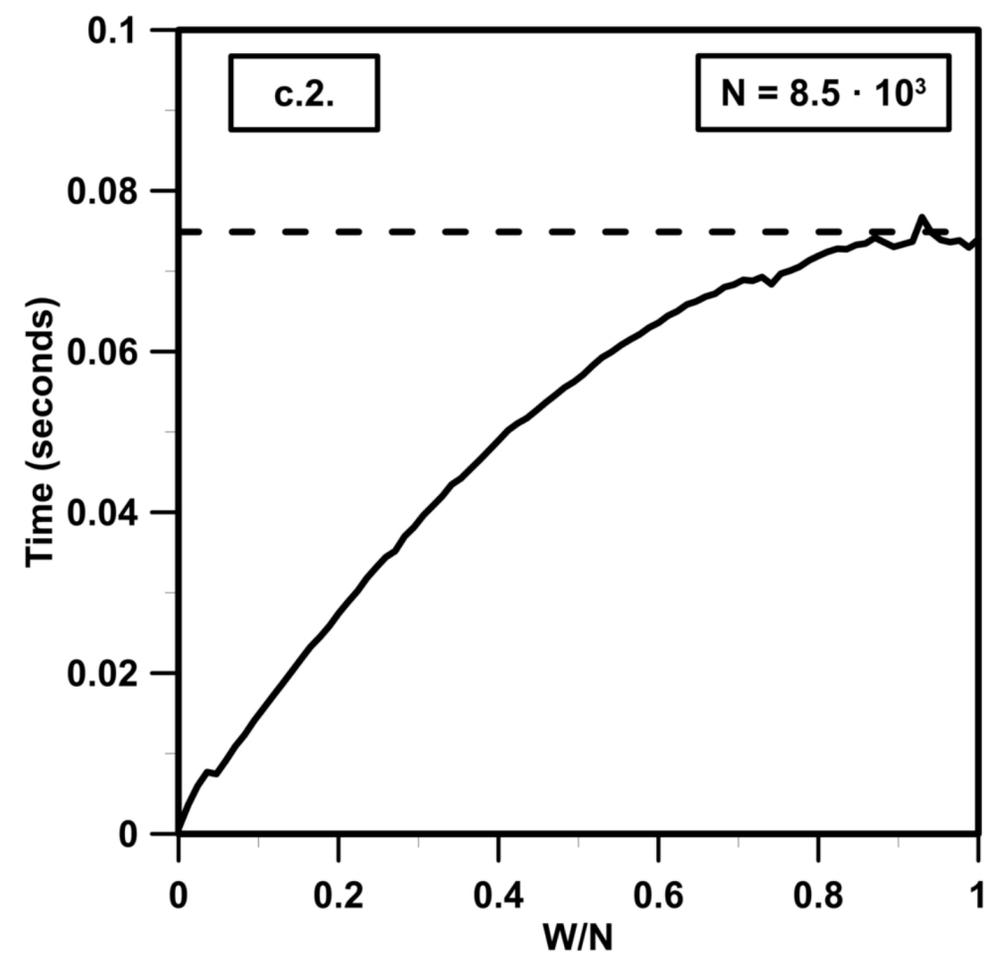

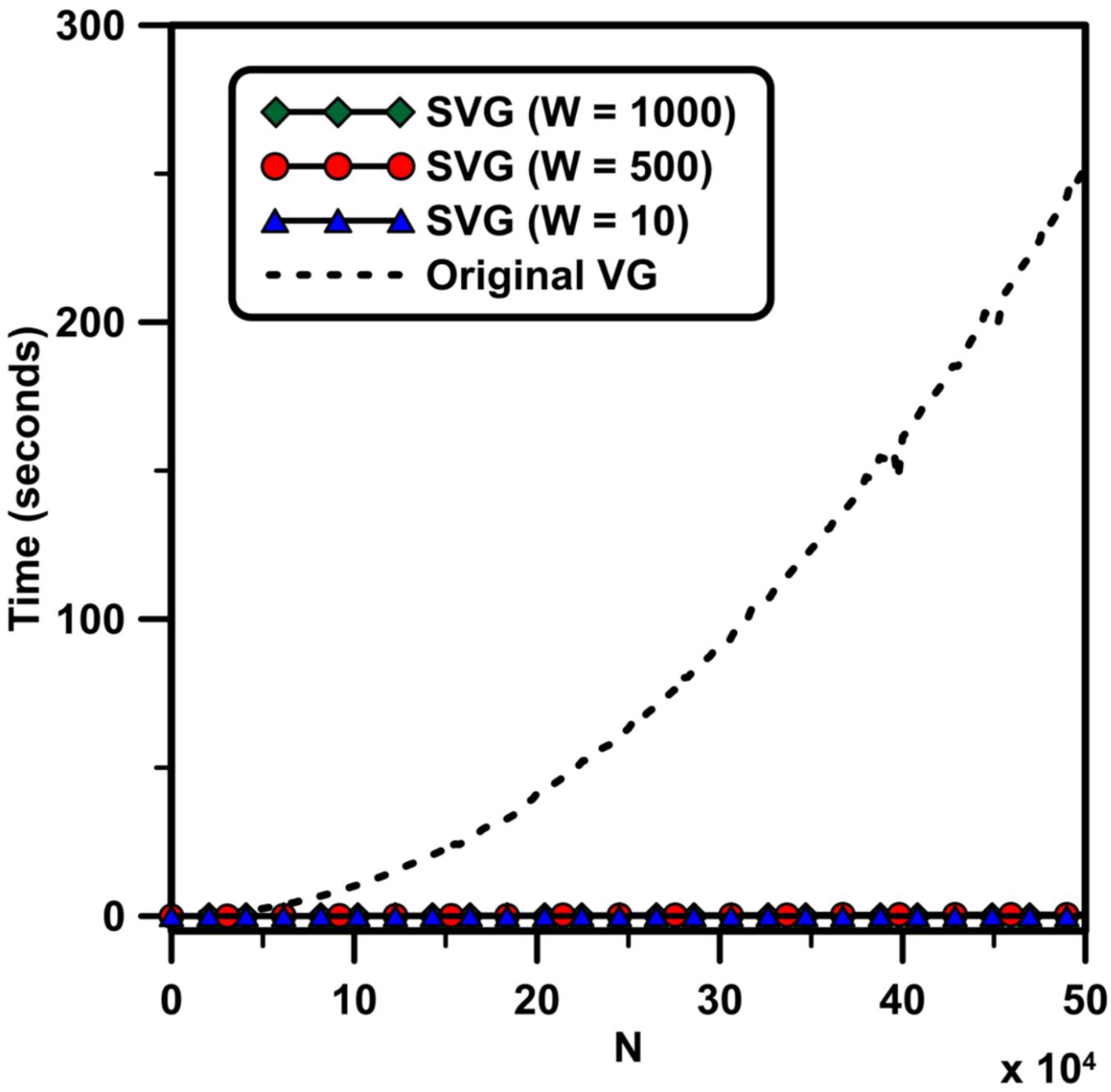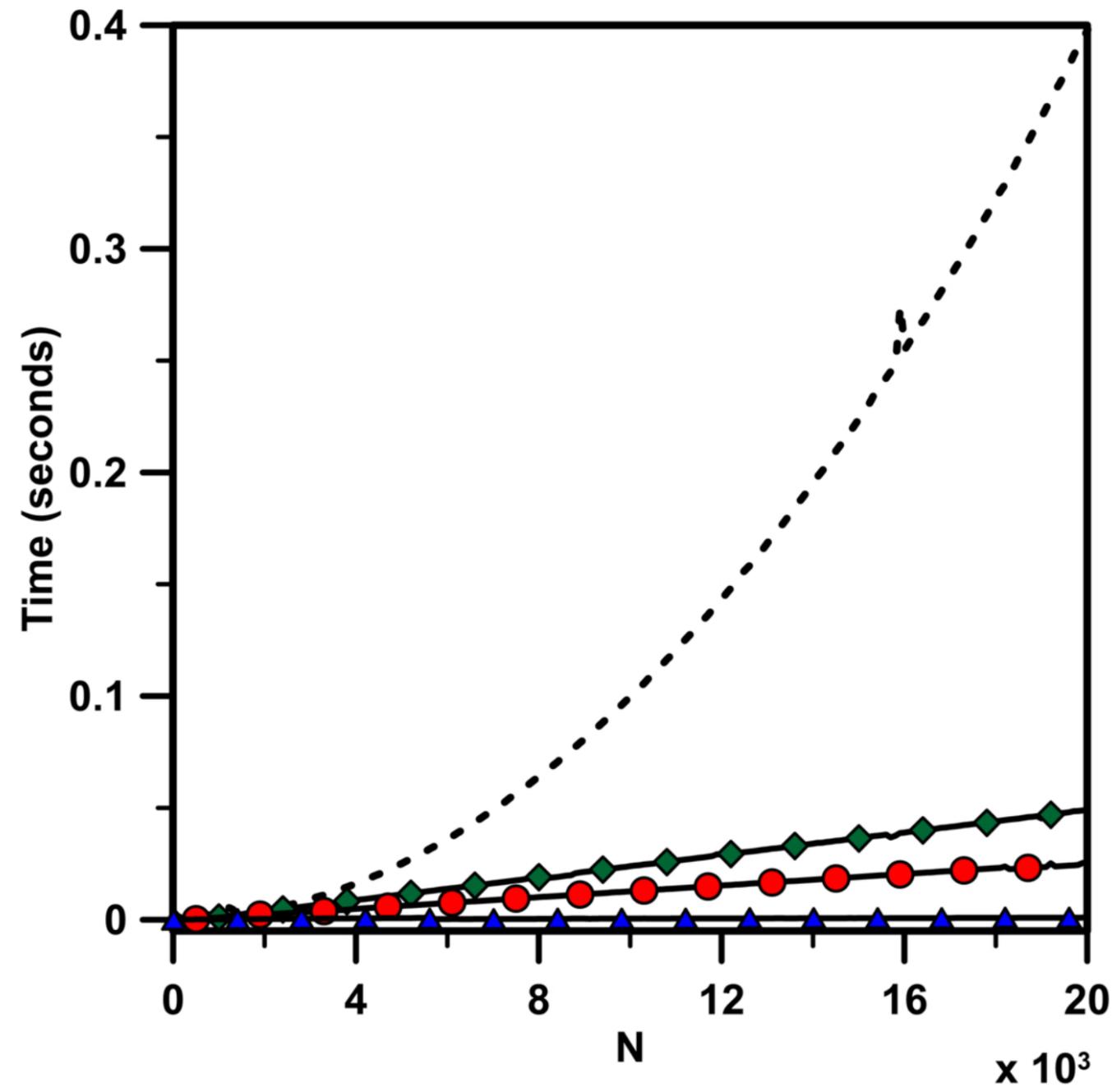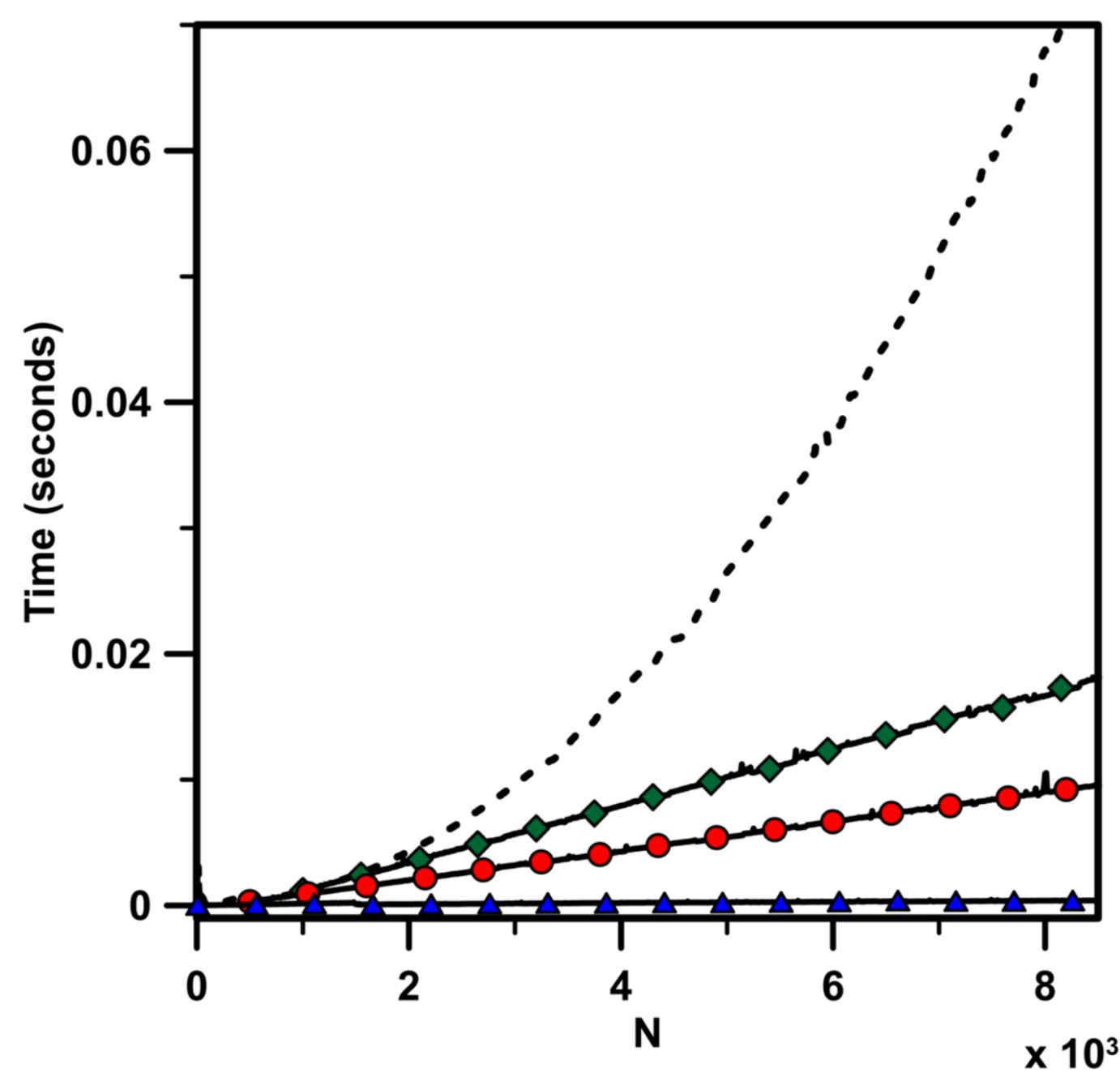

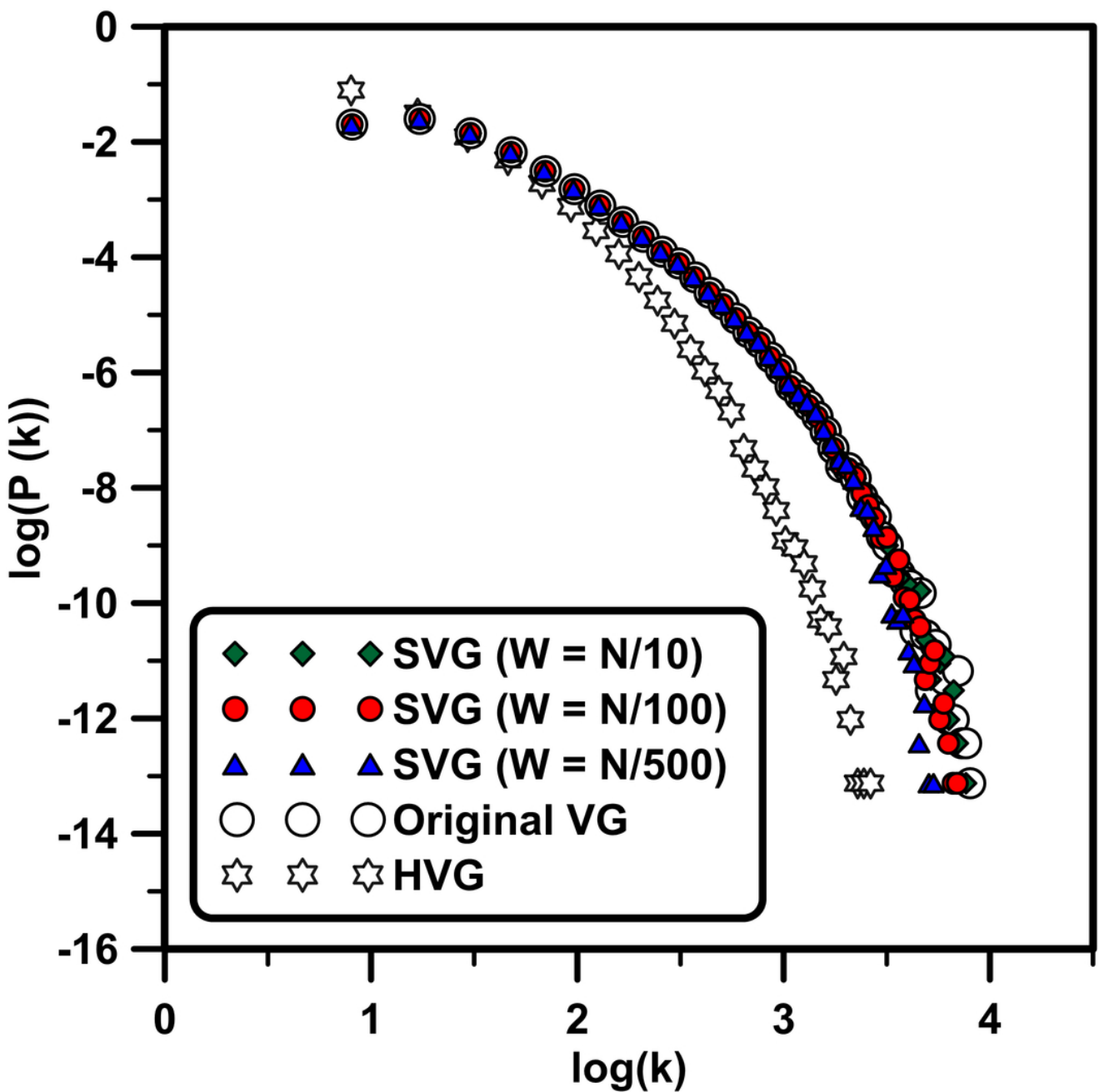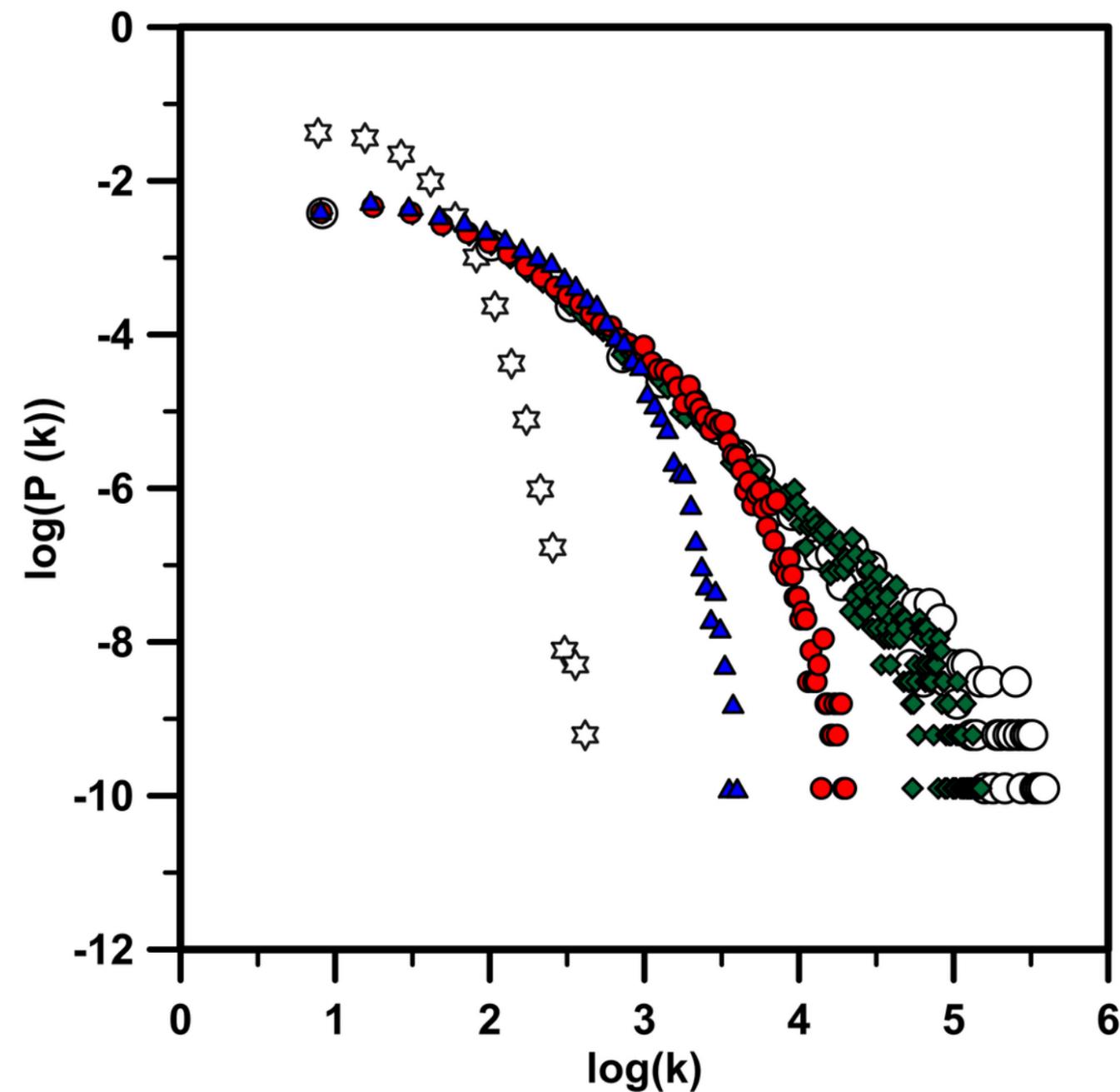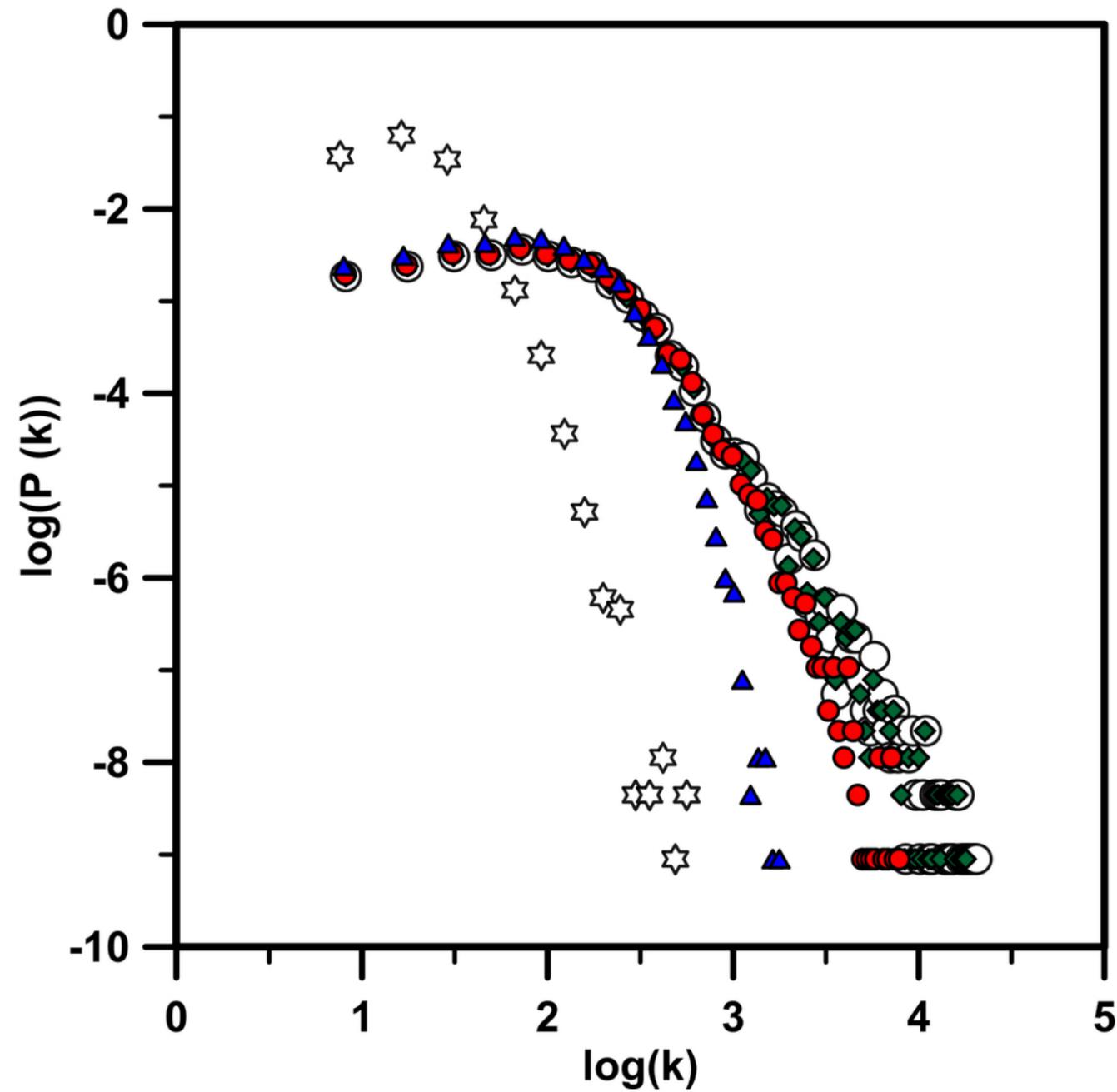

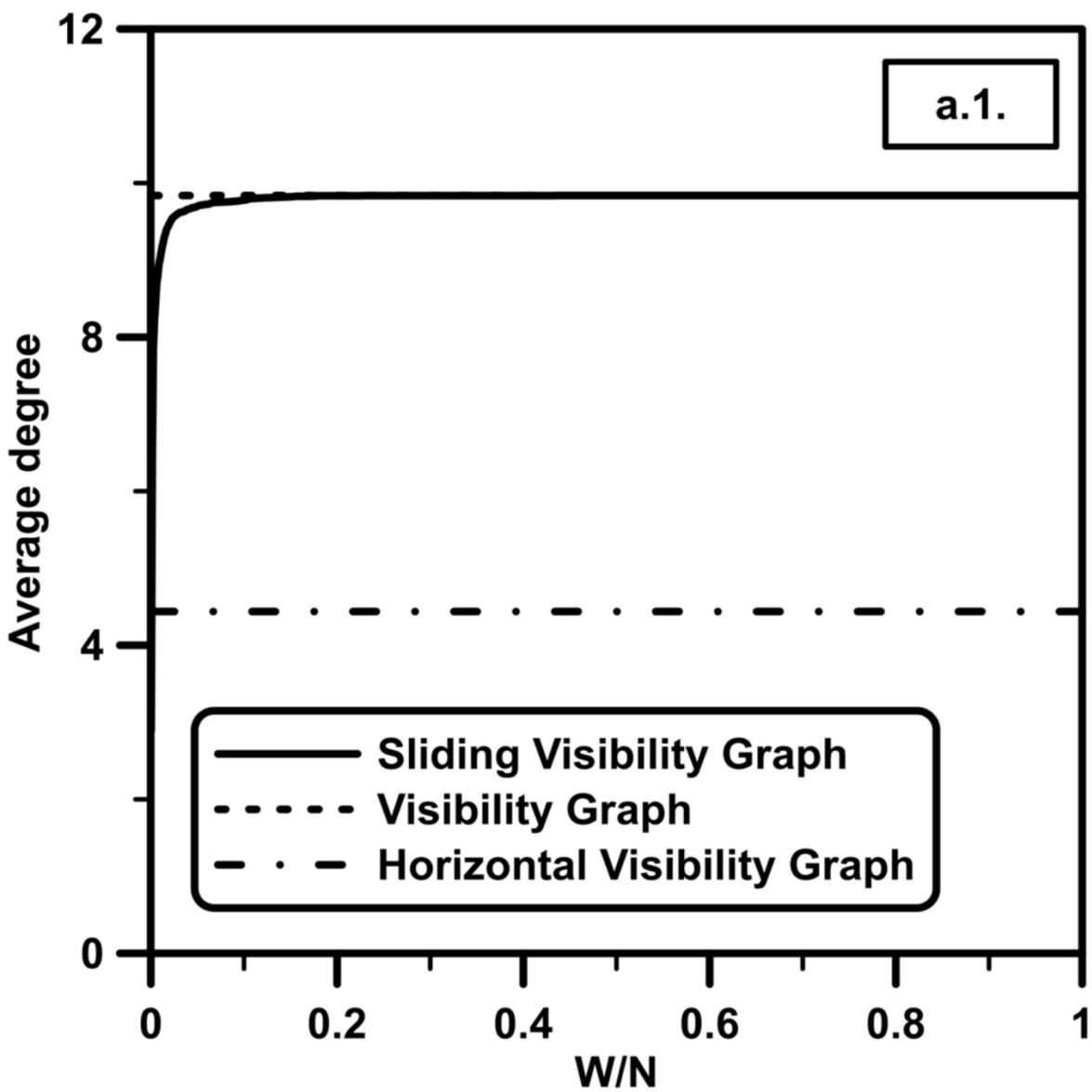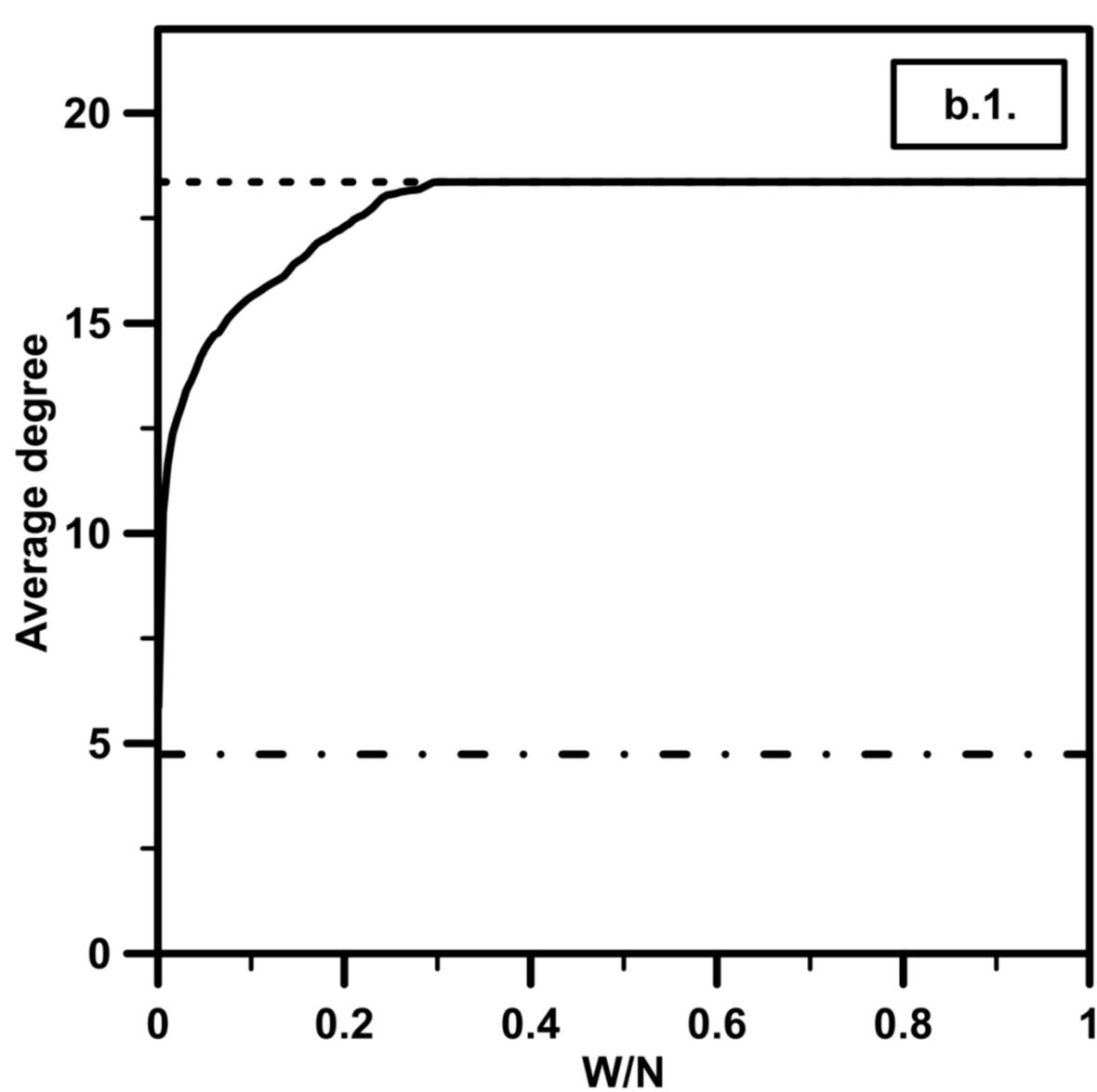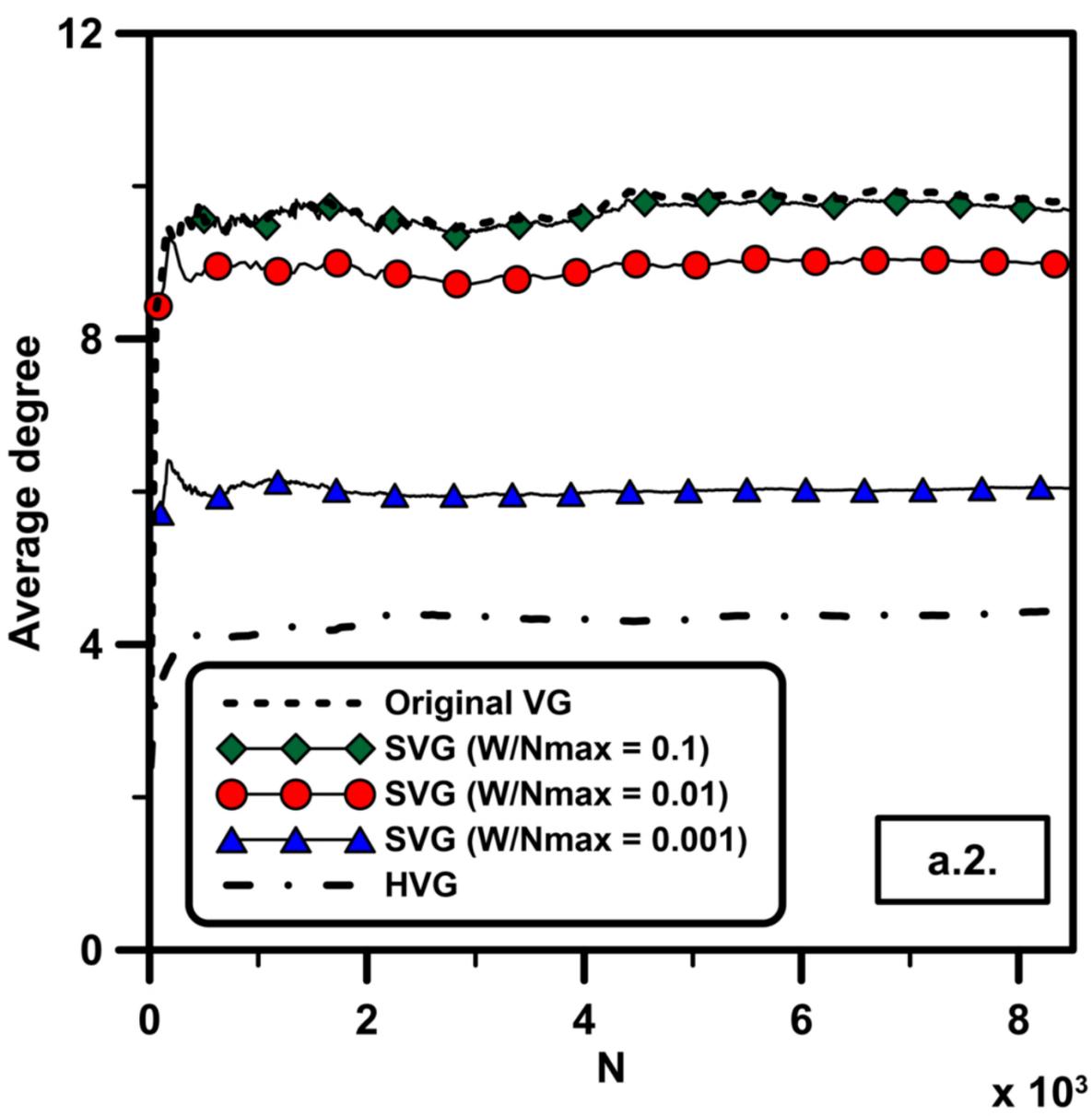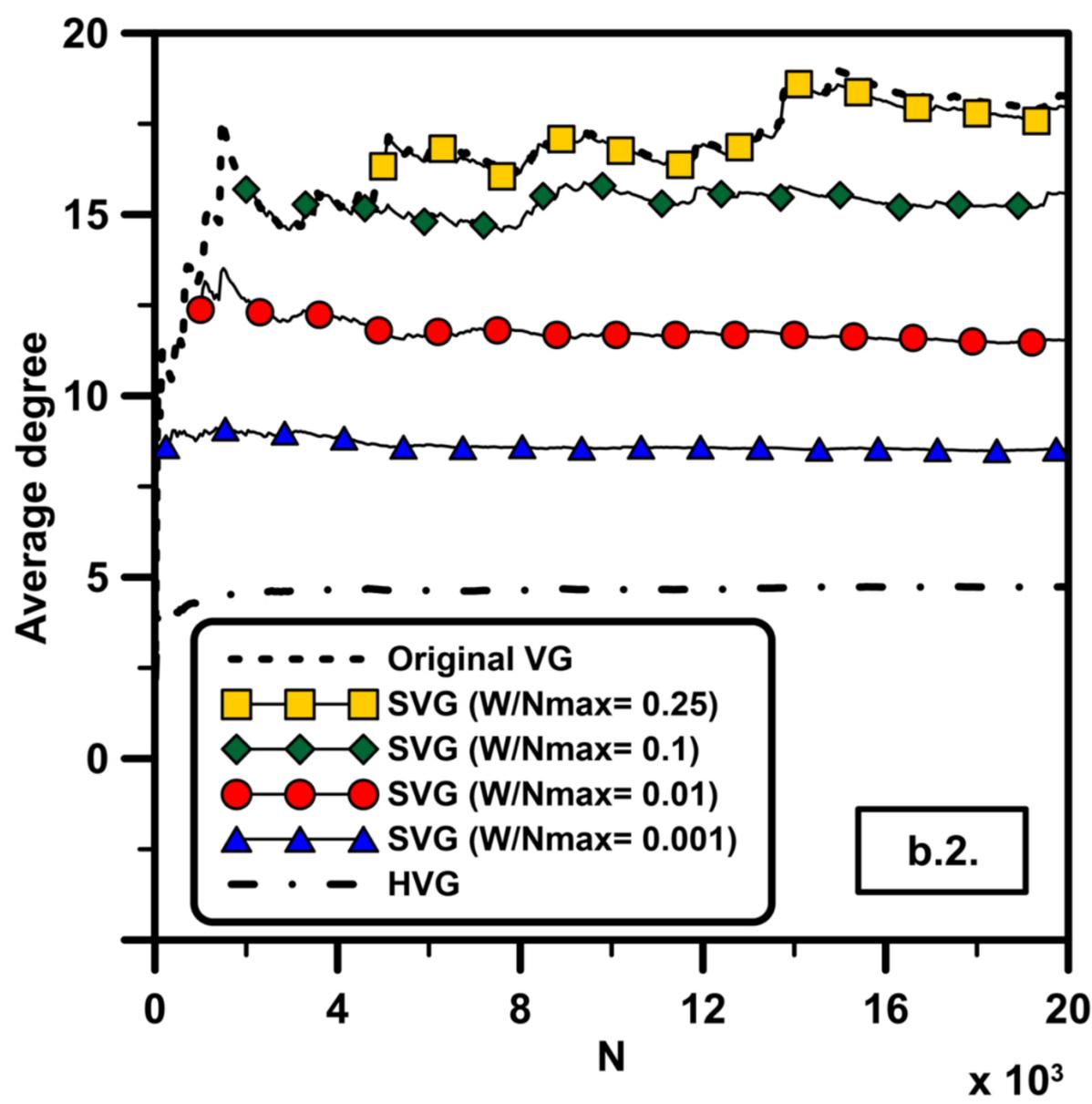

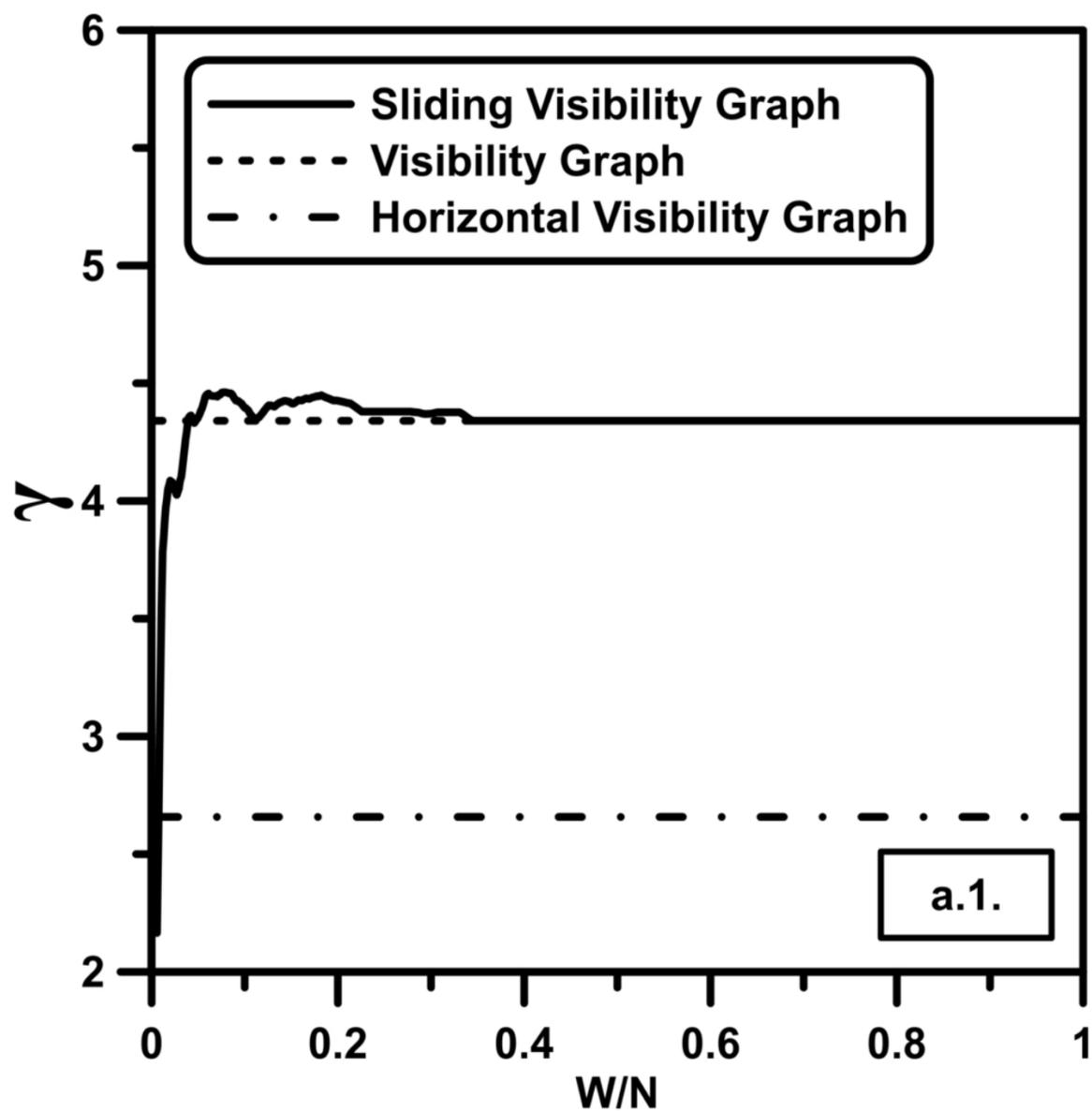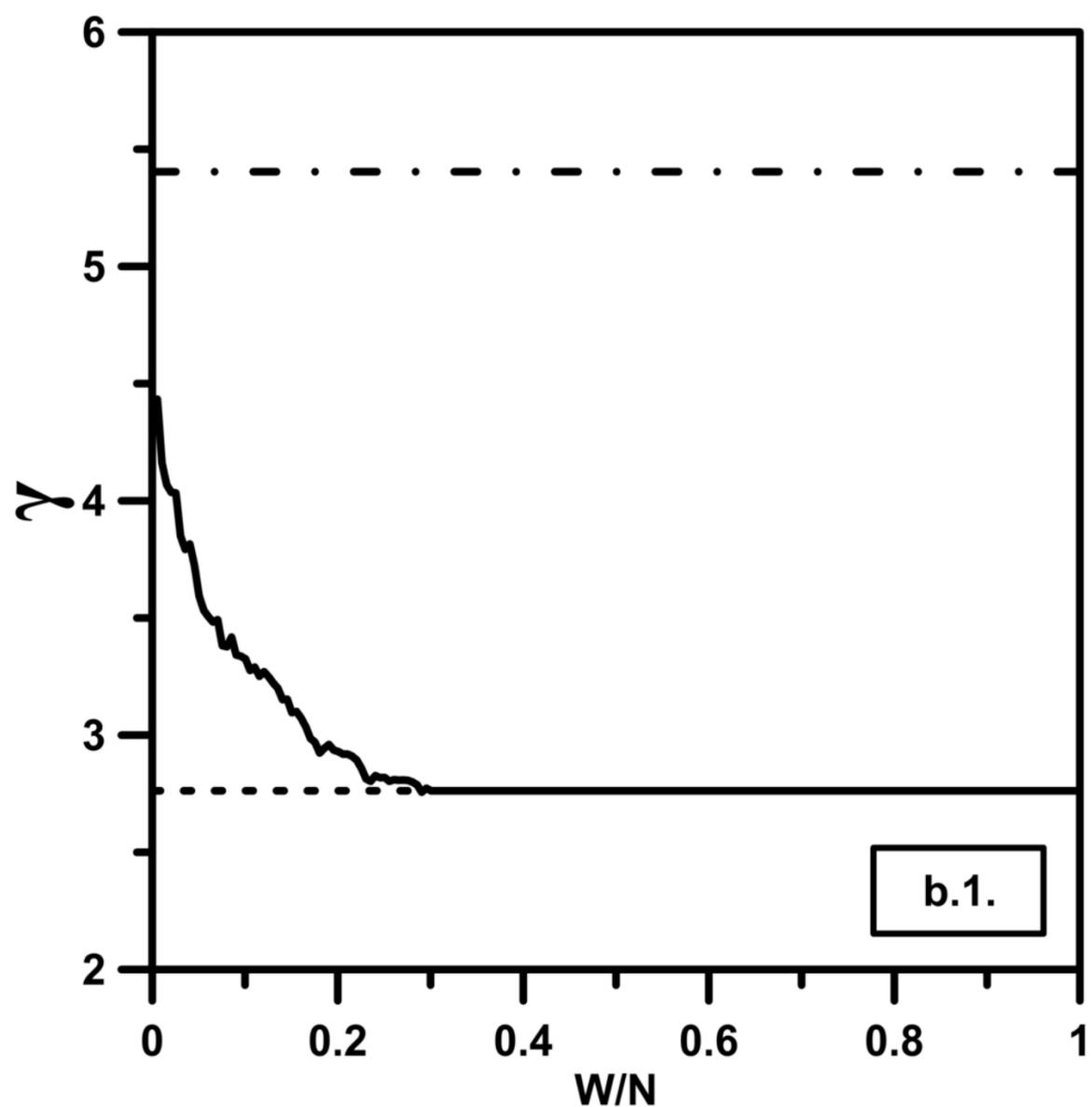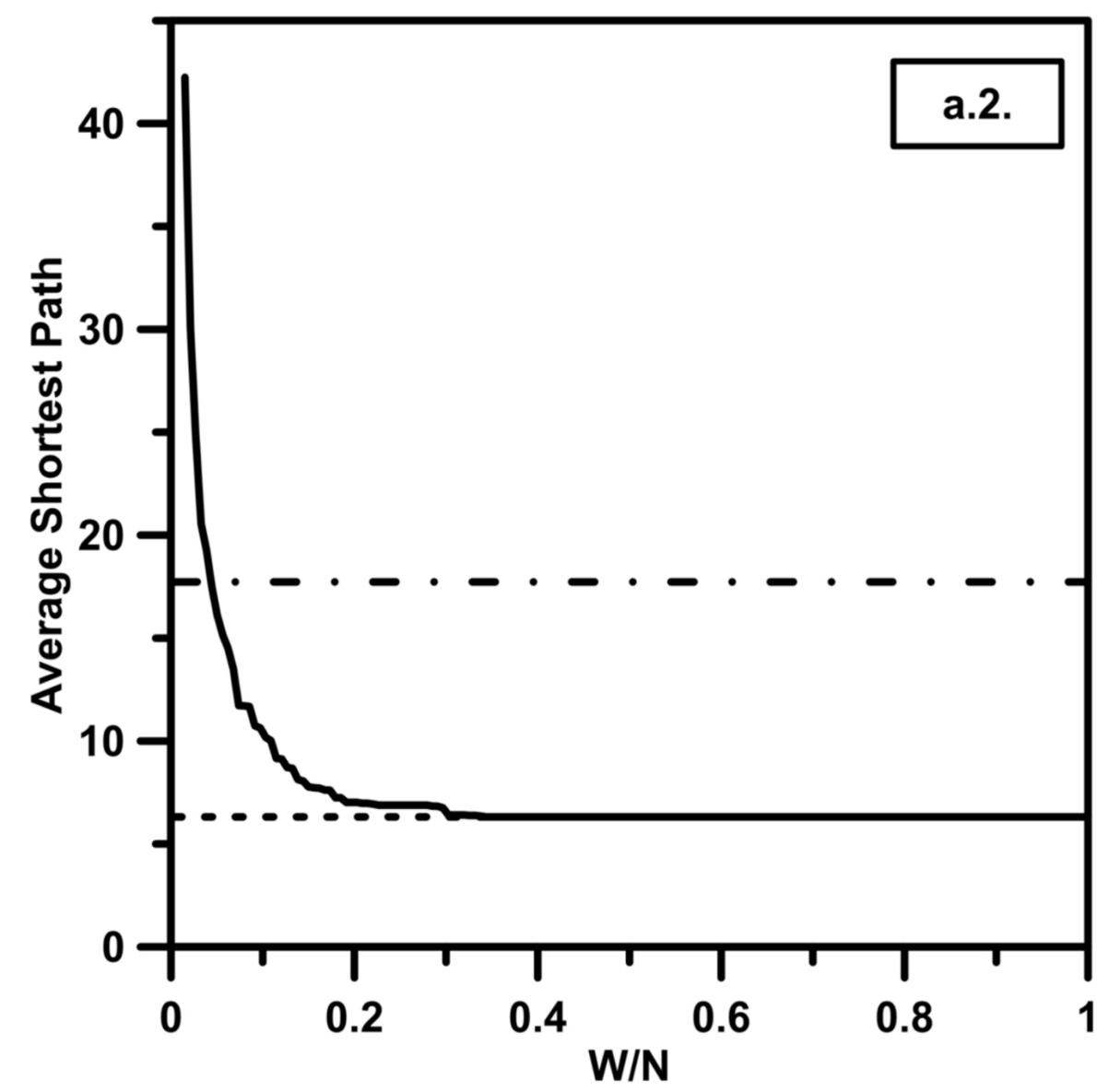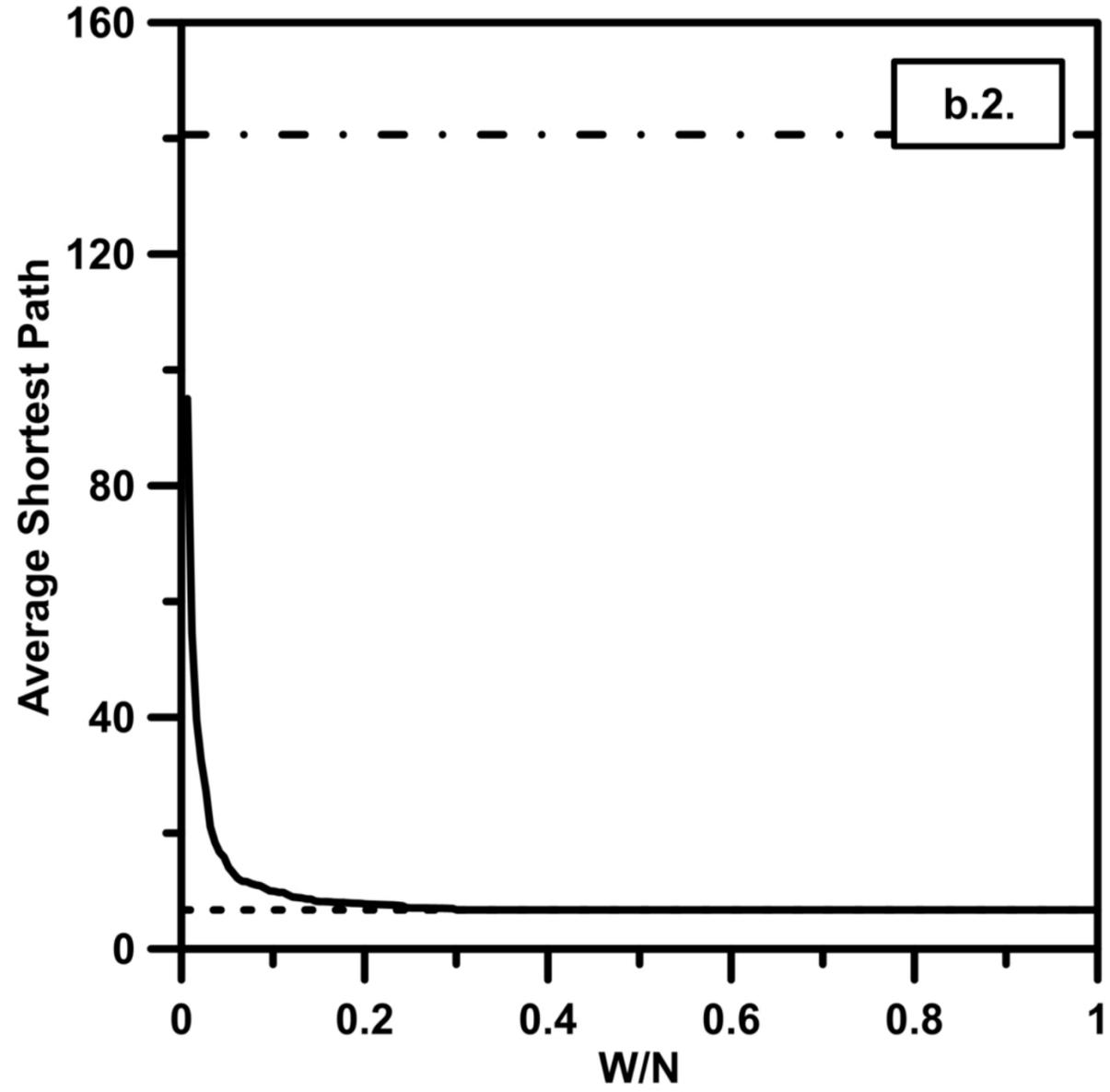

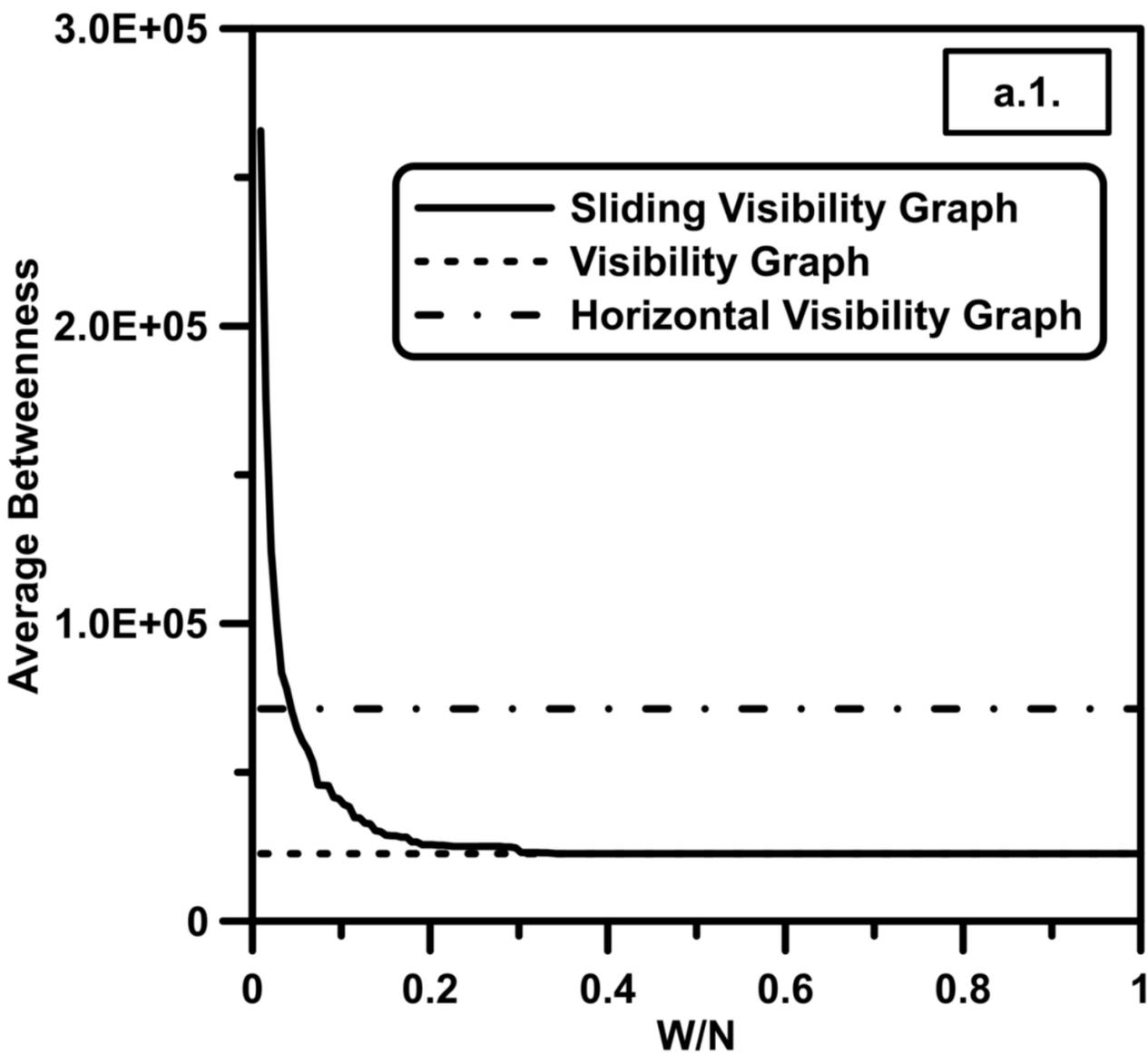
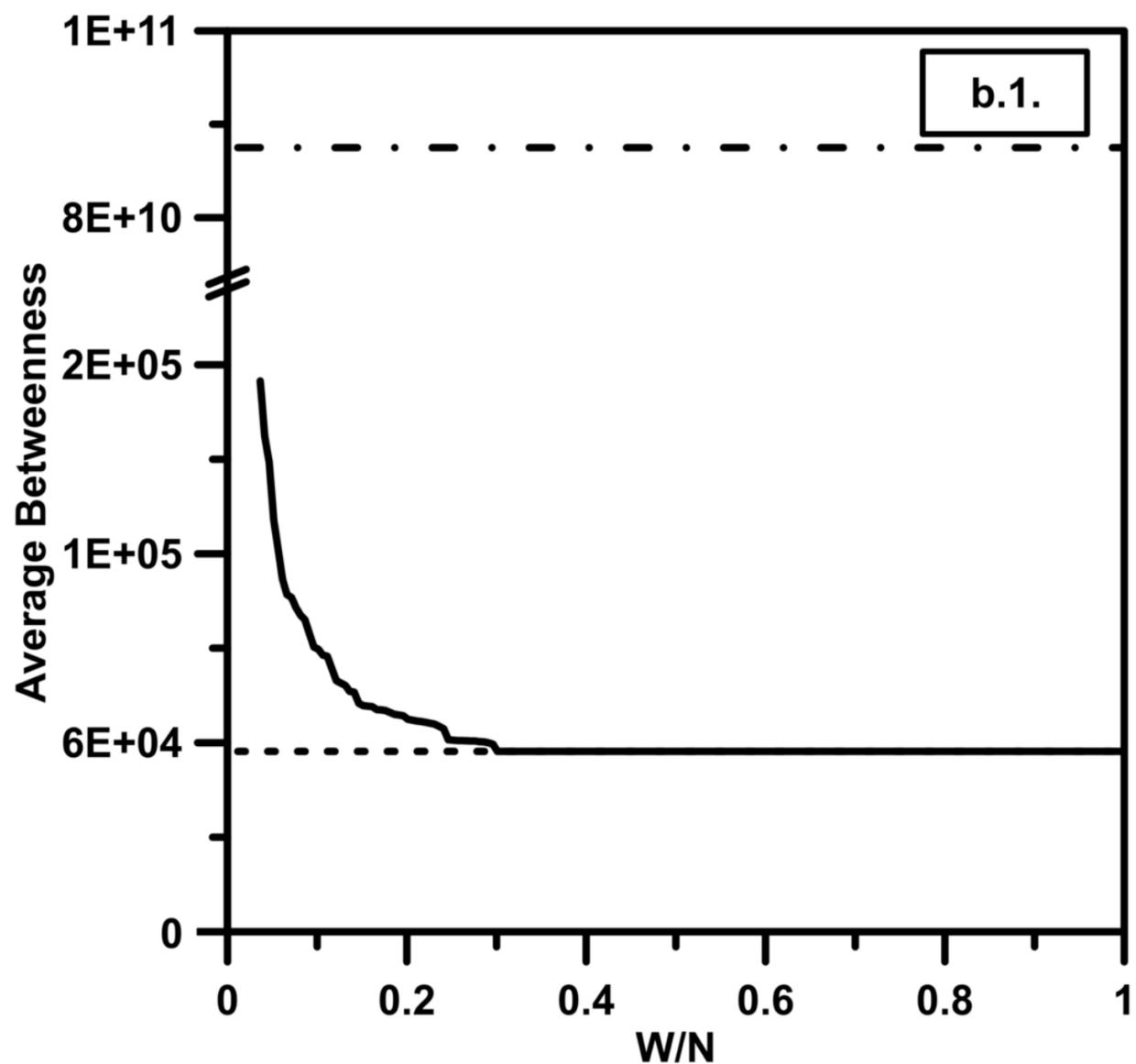
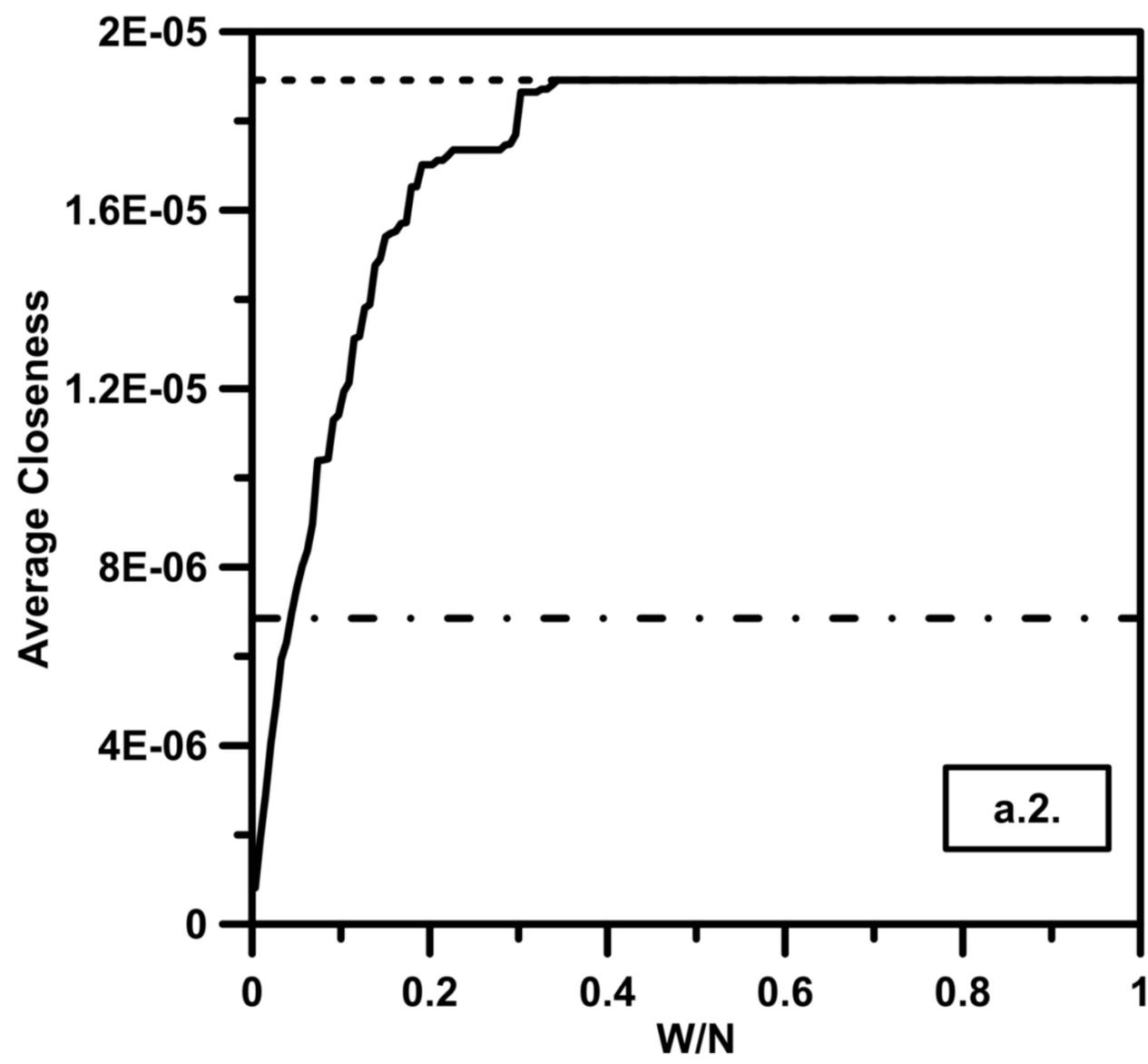
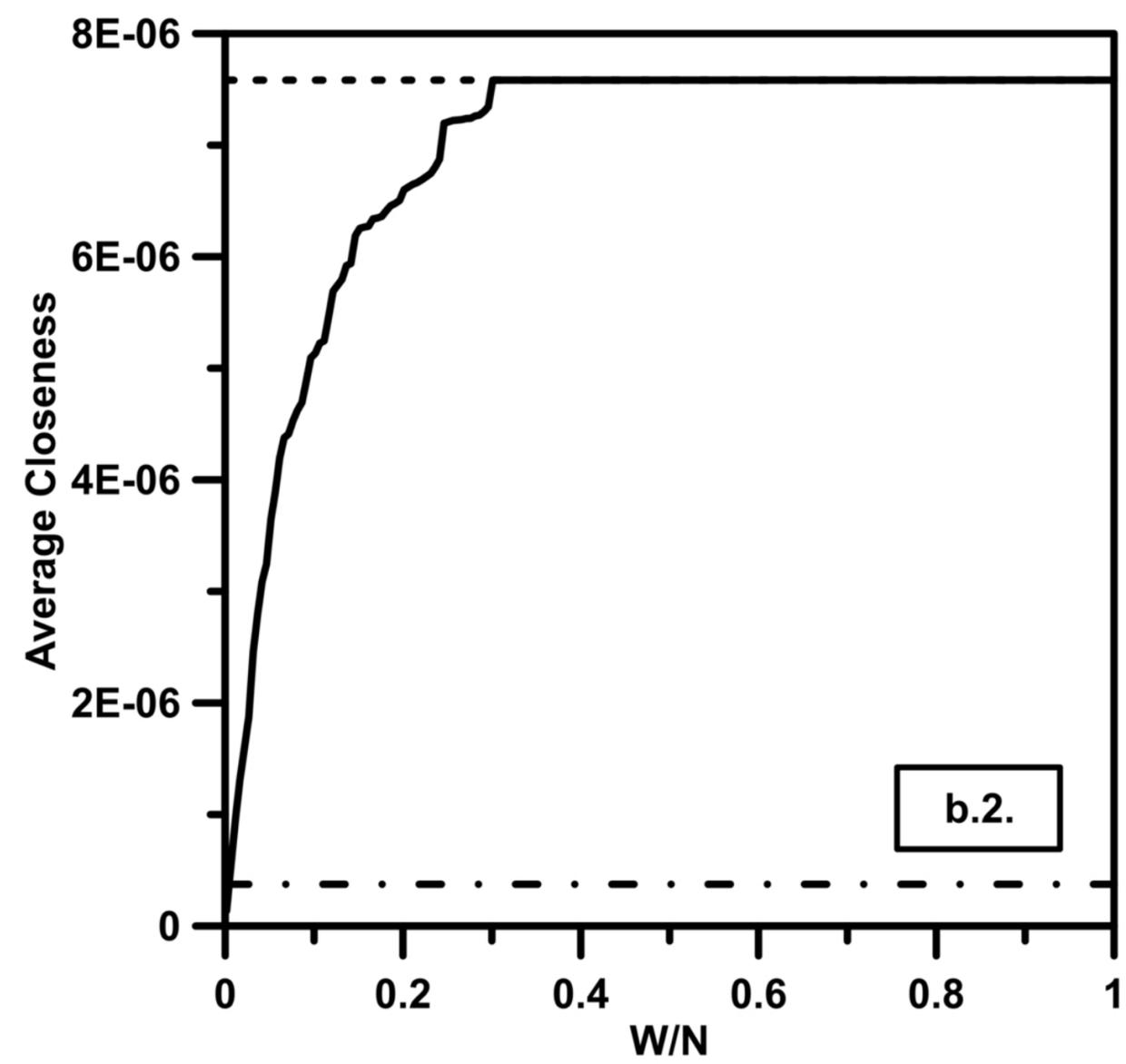